\documentclass[reprint,aps,superscriptaddress,amsmath,amssymb,longbibliography,nofootinbib]{revtex4-1}
\usepackage{graphicx}% Include figure files
\usepackage{dcolumn}% Align table columns on decimal point 
\usepackage{bm}% bold math                                
\usepackage{color}
\usepackage{xspace}
\usepackage{ulem}
\usepackage{xfrac}
\usepackage[colorlinks=true,urlcolor=blue,citecolor=blue,linkcolor=blue,bookmarks=false,pdfstartview={FitH}]{hyperref}

\def \cm-1{cm$^{-1}$\,}
\footnotetext{These authors contributed equally to this work}
\newcommand{\pll}{\kern 0.56em/\kern -0.8em /\kern 0.56em} 
                                                  
\begin{document} 
\title{\textcolor{black}{Temperature-driven sodium-ion dynamical-to-static crossover in the zig-zag ordered phase of Na$_{0.5}$CoO$_2$}}                           

\author{Shangfei~Wu$^\star$}
\email{wusf@baqis.ac.cn}
\affiliation{Beijing Academy of Quantum Information Sciences, Beijing 100193, China}

\author{Hengxin~Tan$^\star$}
\affiliation{Key Laboratory of Artificial Structures and Quantum Control (Ministry of Education), School of Physics and Astronomy, Shanghai Jiao Tong University, Shanghai 200240, China}

\author{Dong~Wu}
\affiliation{Beijing Academy of Quantum Information Sciences, Beijing 100193, China}

\author{Mingshu~Tan}
\affiliation{School of Physical Science and Technology, Lanzhou University, Lanzhou 730000, China}

\author{Xinyu~Zhou}
\affiliation{International Center for Quantum Materials, School of Physics, Peking University, Beijing 100871, China}
   
\author{Tianchen~Hu}
\affiliation{International Center for Quantum Materials, School of Physics, Peking University, Beijing 100871, China}
    
\author{Tao~Dong}
\affiliation{International Center for Quantum Materials, School of Physics, Peking University, Beijing 100871, China}

\author{Feng~Jin}
\affiliation{Beijing National Laboratory for Condensed Matter Physics, Institute of Physics, Chinese Academy of Sciences, Beijing 100190, China}

\author{Qingming~Zhang}
%\email{qmzhang@iphy.ac.cn}
\affiliation{Beijing National Laboratory for Condensed Matter Physics, Institute of Physics, Chinese Academy of Sciences, Beijing 100190, China}

\author{Nanlin~Wang}
%\email{nlwang@pku.edu.cn}
\affiliation{Beijing Academy of Quantum Information Sciences, Beijing 100193, China}
\affiliation{International Center for Quantum Materials, School of Physics, Peking University, Beijing 100871, China}
\affiliation{Tsung-Dao Lee Institute, Shanghai Jiao Tong University, Shanghai 200240, China}

\date{\today}               
                                                                                                                                                                                                              
\begin{abstract}  
     
%\textcolor{black}{The sodium cobaltate Na$_x$CoO$_2$ has been extensively studied for its diverse functional properties along with its rich charge-spin-orbital ordered states. However, less is known about the sodium-ion lattice dynamics.}
We employ polarization-resolved Raman spectroscopy combined with first-principles calculations to study the sodium-ion lattice dynamics in a sodium zig-zag ordered cobaltate compound Na$_{0.5}$CoO$_2$.
We detect two sodium phonon modes for the first time, and their mode frequencies are consistent with first-principles phonon calculations based on an orthorhombic unit cell.
We find that they appear below around \textcolor{black}{$T^*\sim300\pm50$\,K} with large linewidth broadening, much lower than the sodium zig-zag ordering temperature $T_\text{S}\sim460$\,K, and then narrow at lower temperatures. We interpret the sodium-phonon anomalies occurring at $T^*$ as a dynamical-to-\textcolor{black}{static} crossover involving mainly the motion of sodium ions. Our results suggest that the gradual freezing of the sodium ions and the well-defined \textcolor{black}{static} sodium-zigzag order below $T^*$ set the stage for the emergent electronic and magnetic orders in the CoO$_2$ layer of Na$_{0.5}$CoO$_2$.
                                                                       
\end{abstract}                  
                                                                                                                                                                                      
                                                                                                                                                                                                                                                                                                                                                                                         
\maketitle

\section{INTRODUCTION}\label{INTRODUCTION}

The sodium cobaltate Na$_x$CoO$_2$, with a crystal structure featuring hexagonal layers of sodium ions sandwiched between planes of edge-shared CoO$_6$ octahedra, has been widely studied in the past 20 years for its diverse functional properties and rich phase diagram.
These include its use as a cathode, thermoelectric, and superconducting material, as well as the rich charge-spin-orbital ordered states in this system~\cite{Takada2003,Wang_2003_Nature,Foo_2004_PhysRevLett,Huang2004PhysRevB,Roger_2007_nature,Guo_2008,Cao_2012_PhysRevLett,Raveau_2015_review,Wang_2015_batteries,Gilmutdinov_2020_PhysRevMaterials,Forslund_2025_PhysRevResearch,Lichen_Wu100217}. 
Due to the high mobility of the sodium ions and good electronic conductivity, the Na$_x$CoO$_2$ system has the potential to replace lithium-ion batteries for its comparable electrochemical properties and natural abundance.
The diffusive and mobile nature of sodium ions in Na$_x$CoO$_2$ has been studied by
various experimental techniques, such as neutron diffraction~\cite{Argyriou_2007PhysRevB,Medarde_2013_PhysRevLett}, X-ray diffraction~\cite{Willis_2018_Scientific_Reports}, nuclear magnetic resonance~\cite{Gavilano_2004_PhysRevB,Yokoi_2005_JPSJ,Bobroff_2006_PhysRevLett,Lang_2008_PhysRevB.78.155116,Weller_2009_PhysRevLett}, muon spin resonance~\cite{Mansson_2013_Physica_Scripta,Sugiyama_2020_PhysRevB}, quasi-elastic neutron scattering~\cite{Willis_2018_Scientific_Reports,Shah_2021}, and electrochemical measurements~\cite{Shu_2008_PhysRevB,Willis_2018_Scientific_Reports,Ohishi2023}. 
Furthermore, the Na$_x$CoO$_2$ system exhibits various sodium-ordered states that are commensurate to the lattice, depending on the sodium ion content, e.g., $x=0.25, 0.33, 0.43, 0.5, 0.55$, and 0.71~\cite{Zhang_2005_PhysRevB.71.153102,Huang_2004_JPCM,Zandbergen_2004_PhysRevB,Huang_2010_PhysRevLett}.
At these sodium-ordered states, sodium-ion self-diffusion coefficients are significantly smaller than the average diffusion coefficient~\cite{Shu_2008_PhysRevB}.

\begin{figure*}[!ht] 
\begin{center}
\includegraphics[width=2\columnwidth]{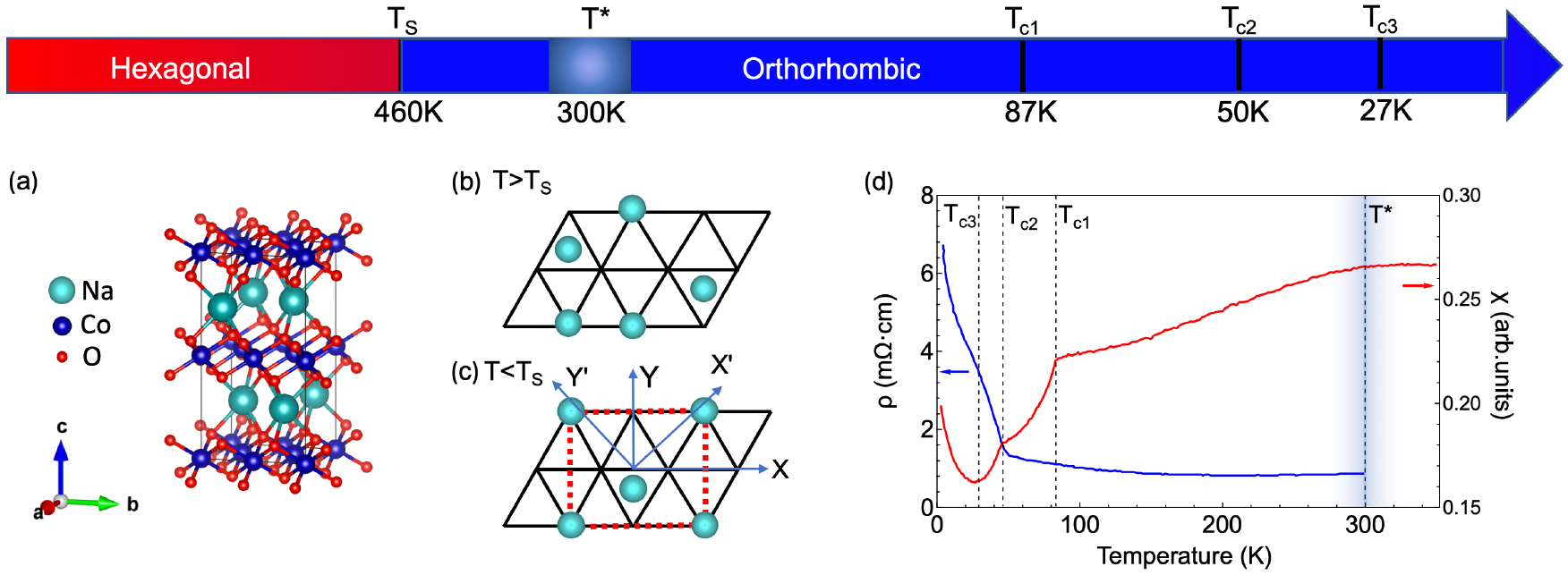}
\end{center}
\caption{\label{Fig1_structure} 
(a) Crystal structure of Na$_{0.5}$CoO$_2$ at 300\,K with space group $Pnmm$ (No.~59) (point group:~$D_{2h}$). 
(b) Na-disordered phase of Na$_{0.5}$CoO$_2$ above $T_\text{S}$.
(c) Na-zigzag-ordered phase of Na$_{0.5}$CoO$_2$ below $T_\text{S}$. 
The triangular lattice represents the Co lattice, and the O atoms are omitted.
The red dashed rectangle in panel (c) denotes the reconstructed $2\times \sqrt 3$ orthorhombic supercell.
The arrows in panel (c) are the definitions of $X$, $Y$, $X'$, and $Y'$ directions. $X'$ and $Y'$ directions are 45 degrees away from the $X$ and $Y$ directions. The $X$ direction is along the hexagonal axis.
(d) Temperature-dependence of the in-plane resistivity data $\rho(T)$, and magnetic susceptibility data $\chi(T)$ with magnetic field 100\,Oe applied parallel to the $ab$-plane. The vertical dashed lines represent magnetic phase transition temperature $T_\text{c1}\sim87$\,K, metal-insulator transition temperature $T_\text{c2}\sim50$\,K, magnetic moment re-ordering temperature $T_\text{c3}\sim27$\,K, and the dynamical-to-\textcolor{black}{static} crossover temperature \textcolor{black}{$T^*\sim 300\pm 50$\,K}.}
\end{figure*}  
   
The $x = 0.5$ compound Na$_{0.5}$CoO$_2$ stands out in the phase diagram of Na$_{x}$CoO$_2$, for its insulating ground state that separates the ‘‘paramagnetic" metallic ground state for $x<0.5$ and the ``Curie-Weiss" metallic ground state for $x>0.5$~\cite{Foo_2004_PhysRevLett,Huang2004PhysRevB}.  As the temperature is varied, 
Na$_{0.5}$CoO$_2$ exhibits several charge-spin-lattice ordered states. Indeed, the sodium ions order into sodium zigzag chains in an orthorhombic superstructure above room temperature at about $T_\text{S}\sim460$\,K~\cite{Argyriou_2007PhysRevB}, leading to two different Co sites~[Figs.~\ref{Fig1_structure}(a), (b), and (c)]. 
The orthorhombic supercell with space group $Pnmm$ is four times the basic hexagonal arrangement, with $a_0=\sqrt3a_H$, $b_0=2a_H$, and $c_0=c_H$ where $a_H$ and $c_H$ are the hexagonal subcell lattice parameters~[Fig.~\ref{Fig1_structure}(c)]~\cite{Huang_2004_JPCM,Zandbergen_2004_PhysRevB}.
At $T_\text{c1}\sim87$\,K, a long-range antiferromagnetic phase transition occurs~\cite{Mendels_2005_PhysRevLett,Gasparovic_2006_PhysRevLett}, associated with a structural distortion~\cite{Williams_2006_PhysRevB}. 
At around $T_\text{c2}\sim50$\,K, a metal-to-insulator transition appears~\cite{Foo_2004_PhysRevLett}, accompanied by an insulating gap of approximately 15\,meV as determined by infrared spectroscopy~\cite{Wang_2004_PhysRevLett}. 
This transition is also regarded as a charge order transition resulting in two Co valence states of  Co$^{3.5+\delta}$ and Co$^{3.5-\delta}$~\cite{Foo_2004_PhysRevLett,Wang_2004_PhysRevLett}.
At a lower temperature, $T_\text{c3}\sim27$\,K, a magnetic moment re-ordering transition emerges~\cite{Yokoi_2005_JPSJ}.
The signatures of these transitions are directly seen from the in-plane resistivity and magnetic susceptibility measurements shown in Fig.~\ref{Fig1_structure}(d). 

\textcolor{black}{Regarding sodium-ion diffusion, the $^{23}$Na NMR study of Na$_{0.5}$CoO$_2$ found that two sodium crystallographic sites can be  resolved only at low temperature ($T < 210$\,K), suggesting sodium-ionic diffusion at higher temperatures~\cite{Bobroff_2006_PhysRevLett}. 
The timescale for sodium-ion self-diffusion in Na$_{0.5}$CoO$_2$ was determined to be microseconds by muon spin resonance~\cite{Mansson_2013_Physica_Scripta,Ohishi2023,tatara2025revisiting}. Electrochemical measurements showed significantly smaller sodium-ion self-diffusion coefficients for Na$_{0.5}$CoO$_2$ and other Na-ordered compounds compared with compounds without Na-order at room temperature~\cite{Shu_2008_PhysRevB,Willis_2018_Scientific_Reports,Ohishi2023,tatara2025revisiting}. 
The polarization-resolved Raman spectroscopy can
probe the sodium lattice degree of freedom in Na$_{0.5}$CoO$_2$, providing a means to study sodium-ion diffusion.} While previous Raman scattering studies on  Na$_{0.5}$CoO$_2$~\cite{ZHANG_2005PhysicaB,Lemmens_2006PhysRevLett,Lemmens_2007PhysRevB,Zhang_2008PhysRevB,Wu_2008_PhysRevB} mainly focus on the Co and O phonon modes, the study of the sodium-ion lattice dynamics and sodium diffusion is not available in the literature.

In this paper, we investigate the sodium lattice dynamics in Na$_{0.5}$CoO$_2$ using polarization-resolved Raman spectroscopy and first-principles calculations. We identify, for the first time, two sodium phonon modes at 129 and 157\,\cm-1, consistent with first-principles phonon calculations based on an orthorhombic unit cell. 
These modes emerge near room temperature (\textcolor{black}{$T^*\sim300\pm50$\,K}) with significant linewidth broadening, well below the established sodium zigzag ordering temperature ($T_\text{S}\sim460$\,K). This anomaly at $T^*$ is interpreted as a crossover from dynamical sodium motion to \textcolor{black}{static} states.
Our findings suggest that the gradual freezing of sodium ions into a well-defined, \textcolor{black}{static} zigzag order below $T^*$ provides the foundation for the emergent electronic and magnetic orders in the CoO$_2$ layer of Na$_{0.5}$CoO$_2$.

\begin{figure*}[t] 
\begin{center}
\includegraphics[width=2\columnwidth]{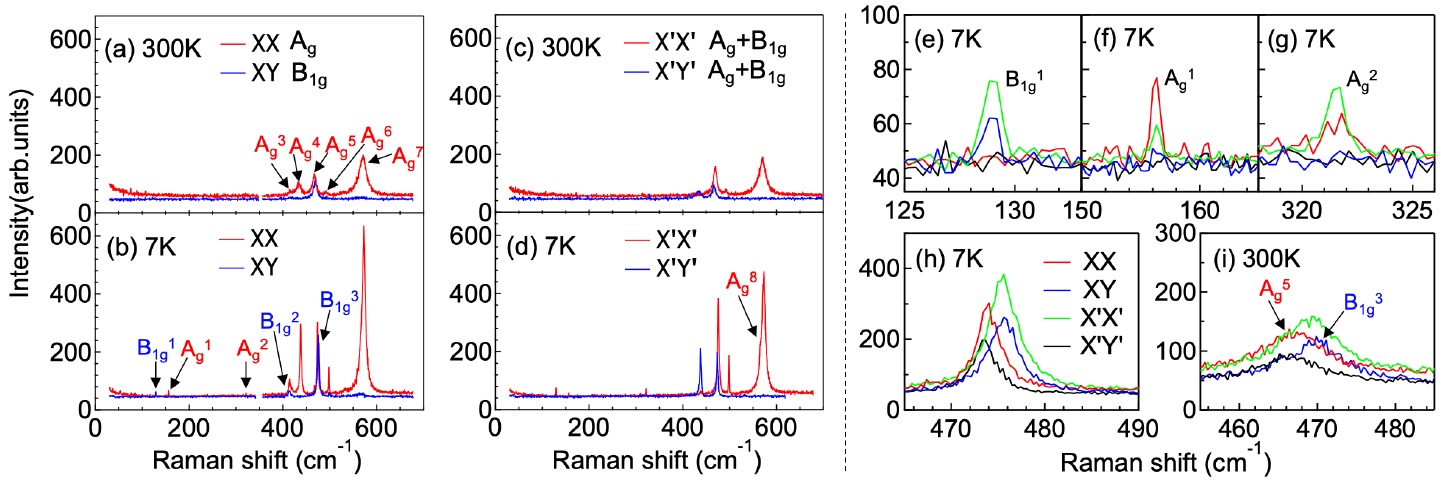}
\end{center}
\caption{\label{Fig2_Raman}
(a)-(b) Raman spectra of Na$_{0.5}$CoO$_2$ at 300\,K and 7\,K in the $XX$ and $XY$ scattering geometries from the $ab$ plane recorded by 633\,nm laser line. The arrows with red labels mark the $A_{g}$ mode, while the arrows with blue labels mark the $B_{1g}$ modes. (c)-(d) same as (a)-(b) but for the $X'X'$ and $X'Y'$ scattering geometries. The arrow in (d) marks the shoulder peak $A^8_{g}$.
(e)-(h) Zoom-in plot of the selected Raman response in the $XX$, $XY$, $X'X'$, and $X'Y'$ scattering geometries at 7\,K. (i) Same as (h) but for 300\,K.}
\end{figure*} 

\begin{figure}[!b] 
\begin{center}
\includegraphics[width=0.9\columnwidth]{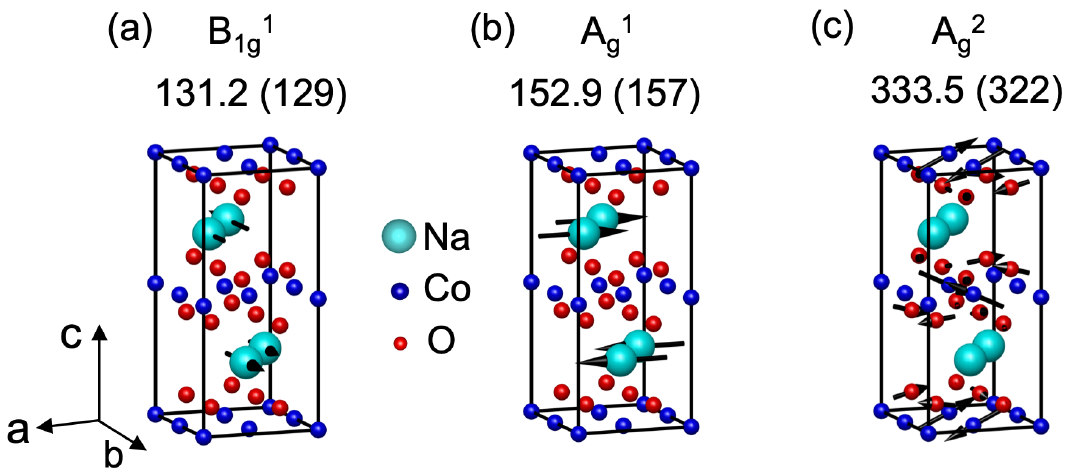}
\end{center}
\caption{\label{Fig3_pattern}
The lattice vibration patterns from the DFT phonon calculations for the first three low-energy modes observed in the experiment. (a) $B_{1g}^1$ mode at 129.8 (129)\,\cm-1. (b)$A_{g}^1$ mode at 154.5 (157)\,\cm-1. (c) $A_{g}^2$ mode at 327.9 (322)\,\cm-1. The corresponding experimental phonon frequencies are in parentheses.
}
\end{figure} 

\section{Methods}\label{Methods}
                                                    
\textit{Single crystal preparation and characterization}\label{Crystal_preparation}
--                                                                                                             
Single crystals of Na$_{0.5}$CoO$_2$ were grown by a floating zone optical image furnace as described in detail in Ref.~\cite{WuDong_2006PRB}. 
Temperature-dependent in-plane resistivity and magnetic susceptibility were measured in a Quantum Design physical property measurement system. A standard four-leads technique was used for resistivity measurement. 
A magnetic field of 100\,Oe was applied parallel to the $ab$-plane in the magnetic susceptibility measurement. The data presented in Fig.~\ref{Fig1_structure}(d) are similar to those in early reports~\cite{Wang_2004_PhysRevLett,Yokoi_2005_JPSJ,Gasparovic_2006_PhysRevLett,WuDong_2006PRB}.
The extracted magnetic phase transition temperature, metal-to-insulator transition temperature, and magnetic moment re-ordering temperature are $T_\text{c1}\sim87$\,K, $T_\text{c2}\sim50$\,K, and $T_\text{c3}\sim27$\,K, respectively.
                                                                                        
\textit{Raman scattering measurements}\label{Raman}
 --                                            
The Na$_{0.5}$CoO$_2$ single crystal is cleaved to expose its (0 0 1) crystallographic planes. Raman scattering experiments were performed with a Horiba Jobin-Yvon LabRAM HR evolution spectrometer.  Volume Bragg gratings were used in the collection optical path to clean the laser line from the backscattered light.
The Na$_{0.5}$CoO$_2$ sample was positioned in a closed-cycle helium gas exchange cryostat (AttoDRY 2100), which allows for cooling down to the base temperature of 1.8\,K.
The Raman measurements were mainly performed using a HeNe laser line at 632.8\,nm (1.96\,eV) in
a backscattering geometry. 
The excitation laser beam was focused into a spot of 5\,$\mu$m diameter, with an incident power of less than 0.5\,mW. 
Linear polarizations were used in this study to distinguish the symmetry of the Raman modes.
%The leakage signals about 7\% for the circular polarization from the optical elements were removed in our data analysis following the method shown in Ref.~\cite{SFWU_PhysRevB2022}.  
Spectra were recorded with a 1800-mm$^{-1}$ grating and a liquid-nitrogen-cooled CCD detector.
The instrumental resolution was maintained better than 0.75\,\cm-1 (full-width at half maximum).
All linewidth data presented were corrected for the instrumental resolution. 
The temperatures shown in this paper were corrected for laser heating with $\sim$10\,K/mW.

%All spectra shown were corrected for the spectral response of the spectrometer and charge-coupled detector to obtain the Raman intensity $I
%_{\mu v}$, which is related to the Raman response $\chi''(\omega,T)$: $I_{\mu v}(\omega, T)=[1+n(\omega, T)] \chi_{\mu \nu}^{\prime \prime}(\omega, T)$. Here, $\mu (v)$ denotes the polarization of the 
%incident (scattered) photon, $\omega$ is the energy, $T$ is the temperature, and $n(\omega, T)$ is the Bose factor.
                                                                                                                                                                                                        
The Raman spectra were recorded from the $ab$ (0~0~1) surface for scattering geometries denoted as $\mu v = XX, XY, X'X', X'Y'$, which is short for $Z(\mu v)\bar{Z}$ in Porto’s notation, where $X$ and $Y$ denote linear polarization parallel and perpendicular to the crystallographic $a$ axis, respectively~[Fig.~\ref{Fig1_structure}(c)]. $X'X'$ and $X'Y'$ denote the directions 45 degrees away from $X$ and $Y$ directions.
The $Z$ direction corresponds to the $c$-axis direction perpendicular to the (0~0~1) plane. 
%Four scattering geometries were employed to probe excitations in different symmetry channels of Na$_{0.5}$CoO$_2$. 

\textit{First-principles calculations}\label{DFT}
 --  
DFT calculations were performed with the Vienna $Ab$-initio Simulation Package (VASP)~\cite{Kresse1996_PhysRevB,KRESSE199615} under the Perdew-Burke-Ernzerhof-type generalized gradient approximation~\cite{Perdew1996PhysRevLett}. 
The projected augmented wave potentials with one valence-electrons for the Na atom, nine valence electrons for Co, and six valence electrons for O were employed. %The cutoff energy for the plane-wave basis set was 400\,eV.  
\textcolor{black}{
Structural relaxations were performed using the conjugate gradient algorithm until the residual forces on each atom were smaller than 5 meV/\AA. The cutoff energy for the plane wave basis set is 400\,eV. The total energy was also converged to an energy threshold of 10$^{-5}$ eV.  A $\Gamma$-centered 9$\times$8$\times$4 $k$-point mesh was used for Brillouin zone sampling. The Gaussian smearing method with a smearing width of 0.1\,eV was adopted. For the phonon calculations using Phonopy~\cite{TOGO20151}, the perturbed structures with distortions were generated with the default displacement amplitude of 0.01\,\AA ~in a 3$\times$2$\times$1 supercell. For self-consistent calculations of these perturbed supercells, the total energy convergence threshold was set to 10$^{-5}$ eV, and a $\Gamma$-centered 2$\times$3$\times$3 $k$-point mesh was used}.         
%The phonon dispersion was calculated by using the finite displacement method as implemented in the phonopy code. 
The on-site Coulomb interaction $U$ of Co $3d$ electrons was set to be 2.5\,eV for the calculations of the band and phonon dispersions~\cite{band_phonon_dispersion}. 
An antiferromagnetic order following Ref.~\cite{Lee_2006_PhysRevLett} was introduced in the calculation.

\textit{Group theoretical analysis}\label{Group}
 --
Group theoretical analysis was performed using the tools provided in the Bilbao Crystallographic Server~\cite{Bilbao_1, Bilbao_4}. The information about the irreducible representations of point groups and space groups follow the notations of Cracknell~$et~al$~\cite{cracknell1979general}.

\section{Results and Discussions}\label{Results} 
          
\subsection{Phonon modes}\label{Phonon}

In Fig.~\ref{Fig1_structure}(d), we present the temperature dependence of the in-plane resistivity data $\rho(T)$, and magnetic susceptibility data $\chi(T)$ with a magnetic field of 100\,Oe applied parallel to the $ab$-plane.
The signatures for the antiferromagnetic phase transition at  $T_\text{c1}\sim87$\,K, metal-insulator transition at $T_\text{c2}\sim50$\,K, and magnetic moment re-ordering transition at $T_\text{c3}\sim27$\,K are clearly seen in the data. Interestingly, we find that the in-plane magnetic susceptibility data is flat above 300\,K, and starts to decrease below 300\,K. The slope of the magnetic susceptibility data vary gradually around 300\,K, suggesting a crossover regime near room temperature. 
This crossover regime is also seen in a previous in-plane magnetic susceptibility data up to 450\,K, which revealed a broad maximum at around $T^*$~\cite{Yokoi_2005_JPSJ}. 

To understand the origin of the crossover regime near $T^*$, we use Raman scattering to study the lattice dynamics of Na$_{0.5}$CoO$_2$. 
Na$_{0.5}$CoO$_2$ belongs to the orthorhombic structure with space group $Pnmm$ (No.~59) (point group:~$D_{2h}$) at room temperature in the presence of the Na zig-zag ordering. 
%The Na1, Na2, Co1, Co2, O1, O2, O3 have Wyckoff positions $2a$, $2b$, $4d$, $4e$, $4e$, $4e$, and $8g$, respectively.  
The Raman tensors $R_{\mu}$ ($\mu=A_{g}, B_{1g}, B_{2g}, B_{3g}$) for the irreducible representations ($\mu$) of point group $D_{2h}$ have the following forms:
\begin{displaymath}   
\left(\begin{array}{ccc}
a & 0  &0\\
0 &b &0\\
0 & 0 &c
\end{array}\right),\\
\left(\begin{array}{ccc}
0 & d  &0\\
e &0 &0\\
0 & 0 &0
\end{array}\right),
\left(\begin{array}{ccc}
0 & 0  &f\\
0 &0 &0\\
g & 0 &0
\end{array}\right),
\left(\begin{array}{ccc}
0 & 0  &0\\
0 &0 &h\\
0 & i &0
\end{array}\right),\\
\end{displaymath}  
respectively. Based on the $D_{2h}$ Raman tensors, the relationship between the scattering geometries and the symmetry channels could be found in Table~\ref{SymmetryAnalysis}.
         
\begin{table}[b]
\caption{\label{SymmetryAnalysis}The relationship between the scattering geometries and the symmetry channels for Na$_{0.5}$CoO$_2$ in the orthorhombic phase. $A_g$ and $B_{1g}$ are the irreducible representations of the $D_{2h}$ point group.}
\begin{ruledtabular}
\begin{tabular}{ccc}
Geometry&Raman tensor elements&Symmetry channels\\
\hline$X X$ & $a^2$ & $A_g$ \\
$Y Y$ & $b^2$ & $A_g$ \\
$X Y$ & $d^2$ & $B_{1g}$ \\
$Y X$ & $e^2$ & $B_{1g}$ \\
$X'X', Y'Y'$ & $(a+b)^2/4+(d+e)^2/4$ & $A_g+B_{1g}$ \\
$X' Y', Y'X'$ & \begin{tabular}{c}
$(a -b)^2/4$+ 
$(d-e)^2/4$
\end{tabular} & $A_g$+$B_{1g}$ \\
\end{tabular}
\end{ruledtabular}
\end{table}

\begin{table}[b]
\caption{\label{phonon_modes} 
Experimental phonon frequencies at the Brillouin zone center for Na$_{0.5}$CoO$_2$ at 7\,K and the phonon frequencies calculated by DFT. All the units are in \cm-1. In Ref.~\cite{ZHANG_2005PhysicaB}, the 475 and 570\,\cm-1 modes are assigned to $E_{1g}$ and $A_{1g}$ modes, respectively.
}
\begin{ruledtabular}
\begin{tabular}{ccccc}
Symmetry ($D_{2h}$) & Cal. &Ref.~\cite{ZHANG_2005PhysicaB} (10\,K)& Exp. (7\,K)&diff.\\
\hline 
$B_{1g}$ & 131.2 && 129 ($B^1_{1g}$)& 2.2\\
$A_g$ & 152.9 & &157 ($A^1_g$)& -4.1\\
$A_g$ & 176.1 && &\\
 $B_{1g}$ & 186.7 &&& \\
 $B_{3g}$ & 230.2 &&& \\
$A_g$ & 250.7 &&& \\
 $B_{3g}$ & 289.4 &&& \\
 $B_{3g}$ & 321.1 &&& \\
$A_g$ & 333.5 & &322 ($A^2_g$)& 11.5\\
 $B_{3g}$ & 342.3 && &\\
 $B_{2g}$ & 381.6 &&& \\
 $B_{1g}$ & 384.4 &414&414 ($B^2_{1g}$)&-29.6 \\
 $B_{3g}$ & 442.6 &&& \\
 $A_g$ & 455.0 &&415  ($A^3_g$)&40\\
$B_{2g}$ & 462.0 & & &\\
 $B_{1g}$ & 462.8 && &\\
 $B_{3g}$ & 467.9 && &\\
$A_g$ & 477.1 &437& 438 ($A^4_g$) &39.1\\
$B_{3g}$& 495.3 && &\\
 $A_g$ & 497.1& 475 ($E_{1g}$)& 474 ($A^5_g$)&23.1\\
 $B_{2g}$ & 500.3 && &\\
 $B_{1g}$ & 504.6 &&476 ($B^3_{1g}$)&28.6\\
 $B_{2g}$ & 513.9 && &\\
 $B_{1g}$ & 516.4 && &\\
 $A_g$ & 519.5 &497&498 ($A^6_g$) &21.5\\
$B_{3g}$ & 519.7 &&  &\\
 $B_{2g}$ & 539.9 &&  &\\
 $B_{1g}$ & 553.4 && &\\
 $A_g$ & 575.2 && 565 ($A^8_g$)&10.2\\
 $B_{3g}$& 575.7 &&  &\\
 $B_{2g}$ & 589.7 && &\\
$A_g$ & 591.0 &570 ($A_{1g}$)& 573 ($A^7_g$)&18\\
 $B_{1g}$ & 592.1 && &\\
  $B_{3g}$ & 596.6 && &\\
 $B_{3g}$ & 599.9 && &\\
$A_g$ & 604.1 && &\\
\end{tabular}
\end{ruledtabular}
\end{table}

From the group theoretical considerations~\cite{Bilbao_1}, $\Gamma$-point phonon modes of the orthorhombic Na$_{0.5}$CoO$_2$ can be expressed as $\Gamma_\text{total}$ = 11$A_{g}$ $\oplus$ 9$A_{u}$ $\oplus$ 6$B_{2g}$ $\oplus$ 14$B_{1u}$ $\oplus$ 8$B_{1g}$ $\oplus$ 14$B_{2u}$ $\oplus$ 11$B_{3g}$ $\oplus$ 11$B_{3u}$. Raman active modes are $\Gamma_{\text{Raman}}$= 11$A_{g}$ $\oplus$ 8$B_{1g}$ $\oplus$ 6$B_{2g}$ $\oplus$  11$B_{3g}$. % IR active modes are $\Gamma_{\text{IR}}$=13$B_{1u}$ $\oplus$ 13$B_{2u}$ $\oplus$ 10$B_{3u}$, the acoustic mode is $\Gamma_{\text{acoustic}}$ =$B_{1u}$ $\oplus$ $B_{2u}$ $\oplus$ $B_{3u}$, and the silent modes are $\Gamma_{\text{silent}}$ = 9$A_{u}$.
Note that
%(1) Co(1) $4d$ sites is not Raman-active while Co(2) $4e$ become Raman-active.
%(2) Each Na(1) and Na(2) site has $A_{g}$  $\oplus$ $B_{1g}$ $\oplus$ $B_{3g}$ Raman-active modes.
$A_g$ and $B_{1g}$ modes are accessible from the $ab$-plane measurement, while $B_{2g}$ and $B_{3g}$ modes are accessible from the $ac$-plane measurement.

Above $T_\text{S}\sim460$\,K, the neutron powder diffraction data indicate the loss of Na zig-zag ordering, and the lattice returns to $P6_3/mmc$ symmetry (space group: No.~194, point group:~$D_{6h}$)~\cite{Argyriou_2007PhysRevB}. %with decompositions of hexagonal Na$_x$CoO$_2$ and Co$_2$O$_3$. 
The Na layer becomes disordered. The two inequivalent Co and O sites in the orthorhombic phase turn into identical in the hexagonal phase, occupying the $2a$ (Co) and $4f$ (O) Wyckoff positions as NaCoO$_2$~\cite{Li_2004_PhysRevB}.
The O sites have Raman active modes $\Gamma_{\text{Raman}}$= $A_{1g}$ $\oplus$ $E_{2g}$ $\oplus$ $E_{1g}$ while the Co sites have no Raman-active modes. The $E_{2g}$ mode is accessible from the $ab$-plane measurement, while the $E_{1g}$ modes is accessible from the $ac$-plane measurement.
Thus, from the group-theoretical analysis, the double degenerate $E_{2g}$(O) mode in the hexagonal phase above $T_S$ splits into the $A_g$ and $B_{1g}$ modes in the orthorhombic phase below $T_S$ from the $ab$-plane measurement.

\begin{figure*}[!t] 
\begin{center}
\includegraphics[width=1.6\columnwidth]{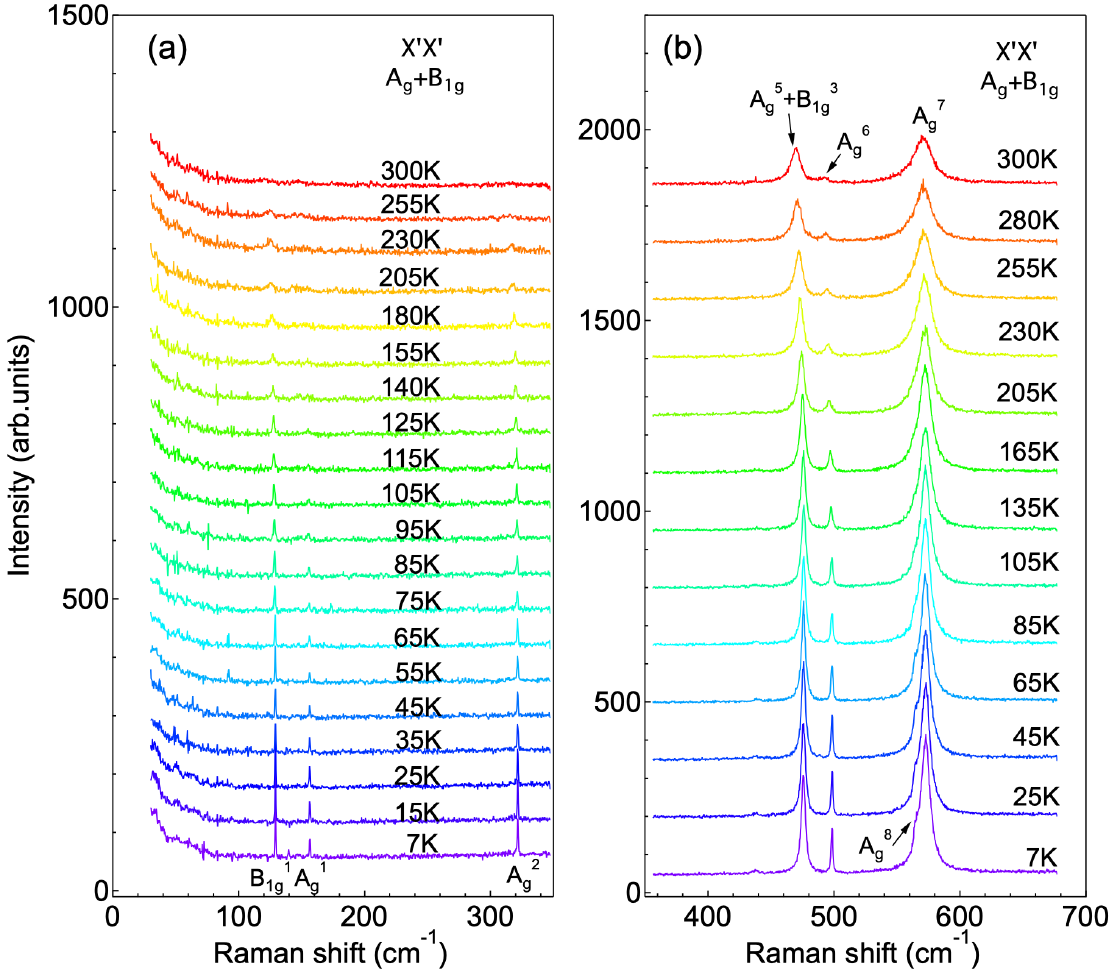}
\end{center}
\caption{\label{Fig4_T_dependence}
Temperature dependence of the Raman spectra for Na$_{0.5}$CoO$_2$ in the $X'X'$ ($A_g$+$B_{1g}$) scattering geometry below 350\,\cm-1 (a) and above 350\,\cm-1 (b).
}
\end{figure*}

In Figs.~\ref{Fig2_Raman}(a) and (b), we show the polarization-resolved Raman response for Na$_{0.5}$CoO$_2$ in the $XX$ and $XY$ scattering geometries at 300\,K and 7\,K. 
At 300\,K, five $A_g$ modes ($A^3_g$, $A^4_g$, $A^5_g$, $A^6_g$, and $A^7_g$) are resolved in the $XX$ scattering geometries. Two $B_{1g}$ modes ($B^2_{1g}$ and $B^3_{1g}$) are resolved in the $XY$ scattering geometries. 
These modes get sharper and stronger at lower temperatures, as we show the data at 7\,K.
Notably, three resolution-limited low-energy modes ($B^1_{1g}$, $A^1_g$, and $A^2_g$ modes) 
emerge only at low temperatures as we show in the 7\,K data in Fig.~\ref{Fig2_Raman}(b). They are absent in the 300\,K data shown in Fig.~\ref{Fig2_Raman}(a). These three modes are also confirmed in the $X'X'$ scattering geometries at 7\,K shown in Fig.~\ref{Fig2_Raman}(d) and they are absent at 300\,K shown in Fig.~\ref{Fig2_Raman}(c). 
The symmetry assignment of these three new modes are confirmed in the polarization dependence in the $XX$, $XY$,  $X'X'$, and $X'Y'$ scattering geometries shown in Figs.~\ref{Fig2_Raman}(e)-(g). 
\textcolor{black}{The observed three new modes are unlikely to be folded modes, as the space group symmetry remains the same below $T_S$~\cite{Huang_2004_JPCM,Williams_2006_PhysRevB,Argyriou_2007PhysRevB}. They are not likely to be defect-activated features, as these modes are super sharp and resolution-limited at low temperatures~[Fig.~\ref{Fig4_T_dependence}(a)].}
Note that a shoulder peak labeled as $A^8_g$ is detected in Fig.~\ref{Fig2_Raman}(d).
The symmetry and frequencies of these modes are summarized in Table~\ref{phonon_modes}. 
As shown in Table~\ref{phonon_modes}, the five modes in the $XX$ scattering geometry, namely, $B^2_{1g}$, $A^4_g$, $A^5_g$, and $A^6_g$ in the frequency range between 400 and 500\,\cm-1, and the $A^7_g$ mode at 573\,\cm-1 are consistent with previous Raman scattering studies~\cite{ZHANG_2005PhysicaB,Lemmens_2006PhysRevLett}.
              
In particular, the frequency of the $A^5_g$ mode at 474\,\cm-1 and the $B^3_{1g}$ mode at 476\,\cm-1 are close to the $E_{2g}$(O) mode of hexagonal Na$_{x}$CoO$_2$~\cite{Lemmens_2006PhysRevLett,Lemmens_2007PhysRevB,Wu_2008_PhysRevB}. These two modes are assigned to be the splitting of the double degenerate $E_{2g}$(O) mode due to the Na zig-zag ordering. This splitting is evident at 7\,K [Fig.~\ref{Fig2_Raman}(h)] as well as 300\,K [Fig.~\ref{Fig2_Raman}(i)]. This is a clear evidence for the orthorhombic phase of Na$_{0.5}$CoO$_2$ from 7\,K up to room temperature 300\,K. This is consistent with previous neutron refinement results that report the $Pnmm$ phase of Na$_{0.5}$CoO$_2$ between 10 and 300\,K~\cite{Huang_2004_JPCM,Williams_2006_PhysRevB,Argyriou_2007PhysRevB}.

To further understand the lattice dynamics of Na$_{0.5}$CoO$_2$, we perform the DFT phonon calculations based on the $Pnmm$ (No.~59) orthorhombic structure at 10\,K~\cite{Huang_2004_JPCM,Williams_2006_PhysRevB,Argyriou_2007PhysRevB}. We summarize the calculated phonon frequencies and the experimental phonon frequencies at the Brillouin zone center in Table~\ref{phonon_modes}. Overall, the experimentally observed phonon frequencies are consistent with the DFT phonon calculations. Especially, the low-energy modes below 350\,\cm-1 are in good accordance with DFT phonon calculations. For the high-energy modes above 350\,\cm-1 which are mainly O-related modes, the calculated phonon frequencies are, on average, about 30\,\cm-1 larger than the experimental ones. 
\textcolor{black}{The possible sources of the large discrepancies for the high-energy modes are discussed in Supplemental Material~\cite{band_phonon_dispersion}.}
Furthermore, the lattice vibration patterns for each of the $A_g$ and $B_{1g}$ Raman-active modes are shown in Supplemental Material~\cite{band_phonon_dispersion}.
To our surprise, the lowest two experimental phonon modes ($B^1_{1g}$ and $A^1_{g}$ modes) are Na-dominant lattice vibration modes. Their lattice vibration patterns are shown in Fig.~\ref{Fig3_pattern}(a) for the $B^1_{1g}$ mode at 129\,\cm-1 and  Fig.~\ref{Fig3_pattern}(b) for the $A^1_{g}$ mode at 157\,\cm-1. The lattice vibration pattern for the $A^2_{g}$ mode at 322\,\cm-1 is shown in Fig.~\ref{Fig3_pattern}(c), which has dominant Co and O lattice vibration displacements.

\subsection{Temperature dependence} \label{Temperature_dependence}

\begin{figure*}[!t] 
\begin{center}
\includegraphics[width=2\columnwidth]{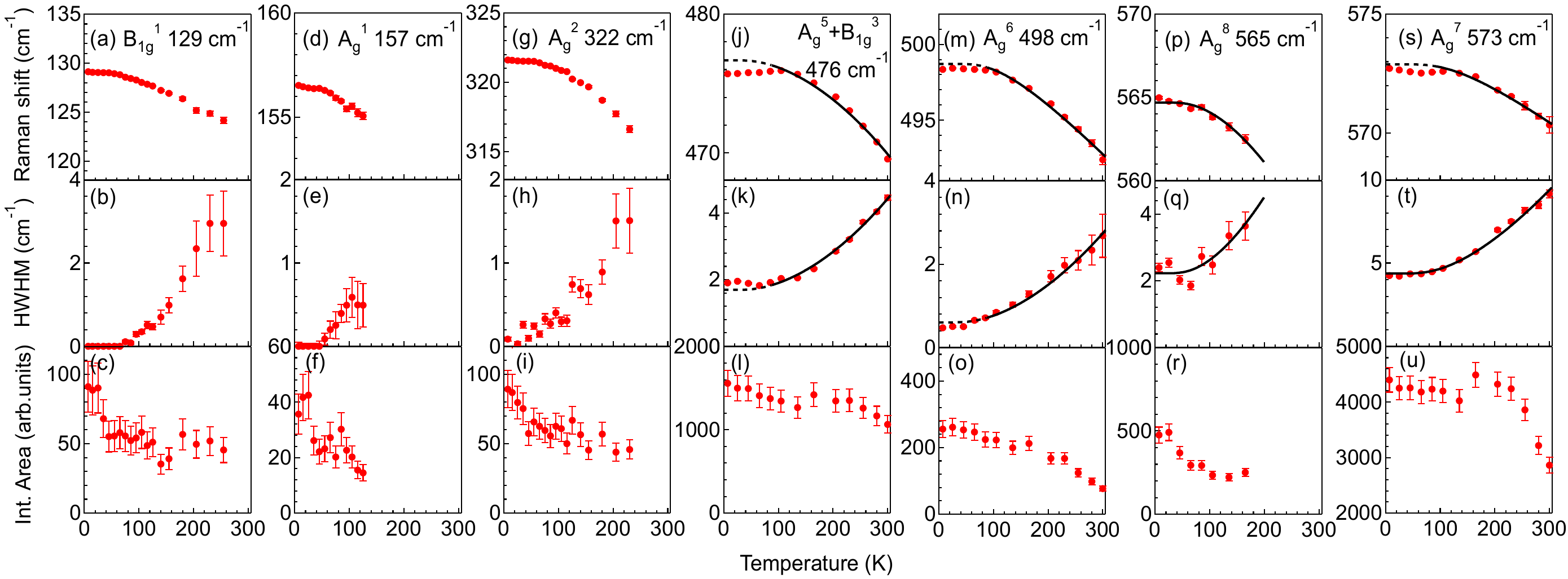}
\end{center}
\caption{\label{Fig5_fitting_parameters}
Temperature dependence of the peak position, HWHM (half-width at half-maximum), and integrated intensity for the Raman modes of Na$_{0.5}$CoO$_2$. (a)-(c) For the $B^1_{1g}$ mode at 129\,\cm-1. (d)-(f) For the $A^1_g$ mode at 157\,\cm-1. (g)-(i) For the $A^2_g$ mode at 322\,\cm-1. (j)-(l) For the mode at around 476\,\cm-1 which contains the $A^5_g$ and $B^3_{1g}$ modes.
 (m)-(o) For the $A^6_g$ mode at 498\,\cm-1. (p)-(r) For the $A^8_g$ mode at 565\,\cm-1. (s)-(u) For the $A^7_g$ mode at 573\,\cm-1. The solid black lines in (j), (k), (m), (n), (p), (q), (s), and (t) are the fitting of the temperature dependence of the peak position and HWHM for the Raman modes above $T_\text{c1}$ according to the anharmonic phonon decay model (Appendix~\ref{Anharmonic_decay_model}). The error bars represent one standard deviation. }
\end{figure*}   

After establishing the Raman-active modes of Na$_{0.5}$CoO$_2$ at 300\,K and 7\,K, we present their temperature dependence in the following.

In Fig.~\ref{Fig4_T_dependence}, we show the temperature dependence of the Raman spectra in the $X'X'$ ($A_g$+$B_{1g}$) scattering geometry. We choose this scattering geometry to study their temperature dependence because all the phonons accessible from the $ab$-plane measurement appear in this scattering geometry, except that the $A^3_g$ and $A^4_g$ modes are very weak to be detected in this scattering geometry.
The phonon modes presented in Fig.~\ref{Fig4_T_dependence} show hardening and narrowing upon cooling. In particular, the two low-energy modes ($B^1_{1g}$ and $A^2_g$) emerge below \textcolor{black}{$T^*\sim300\pm50$\,K}, while the $A^1_g$ mode emerges at a lower temperature at about  $\sim$150\,K~[Fig.~\ref{Fig4_T_dependence}(a)]. 
\textcolor{black}{The large sodium modes’ appearance temperature regime reflects our experimental uncertainties.}
A shoulder peak $A^8_g$ is clearly seen as a function of temperature~[Fig.~\ref{Fig4_T_dependence}(b)].

%Their temperature dependences in the nonmagnetic phase are consistent with the conventional phonon anharmonic decay model~\cite{Klemens_PhysRev148,Cardona_PRB1984}. 
                                                                                                    
In Fig.~\ref{Fig5_fitting_parameters}, we present the temperature dependence of the peak position, HWHM (half-width at half-maximum), and integrated intensity for the Raman modes of Na$_{0.5}$CoO$_2$. 
For the temperature dependence of the peak position, all the modes harden upon cooling. The high-energy modes mainly involving Co or O lattice vibration modes, e.g.,~476, 498, 573\,\cm-1, follow the conventional phonon anharmonic decay model upon cooling to $T_\text{c1}$~\cite{Klemens_PhysRev148,Cardona_PRB1984}. However, they show noticeable softening upon cooling below the antiferromagnetic phase transition temperature $T_\text{c1}$~[Figs.~\ref{Fig5_fitting_parameters}(j), (m), and (s)], suggesting that these lattice vibration modes modulate the Co-Co exchange energy and thus couple to the antiferromagnetic order~\cite{Zhang_2008PhysRevB}. 
In contrast, the Na-lattice vibration modes ($B^1_{1g}$ and $A^1_{g}$) harden smoothly across the magnetic phase transition. They do not show any anomalies upon cooling across $T_\text{c1}$~[Figs.~\ref{Fig5_fitting_parameters}(a) and (d)]. 
This might be because the Na layer is away from the Co layer and Na-lattice vibration modes barely modulate the Co-Co exchange energy, thus do not couple to the antiferromagnetic order.                          

For the temperature dependence of the HWHM, they decrease upon cooling. The high-energy modes at 476, 498, 563, 573\,\cm-1 follow the conventional phonon anharmonic decay model upon cooling to $T_\text{c1}$~\cite{Klemens_PhysRev148,Cardona_PRB1984}. The modes at 476 and 498\,\cm-1 deviate slightly from the anharmonic decay model below $T_\text{c1}$~[Figs.~\ref{Fig5_fitting_parameters}(k) and (n)]. 
For the Na-lattice vibration $B_{1g}^1$ mode at 129\,\cm-1 and $A_g^1$ mode at 157\,\cm-1, their intrinsic HWHMs approach zero because these two modes are resolution-limited below about 60\,K~[Figs.~\ref{Fig5_fitting_parameters}(b) and (e)]. 
These are in contrast to  Co or O lattice vibration modes, which have an intrinsic HWHM of about 1-2\,\cm-1 at 7\,K~[Figs.~\ref{Fig5_fitting_parameters}(k), (n), (q)]. The resolution-limited Na-lattice vibration modes indicate the high quality of the single crystal, and that the Na-lattice is well-ordered with long-range correlation at lower temperatures. This is consistent with the thermal conductivity $\kappa$ measured parallel to the layers that $\kappa$ rises steeply below $T_\text{c2}$ in Na$_x$CoO$_2$ crystals with $x=0.5$, while it is only weakly temperature dependent in metallic samples ($x=0.31$ and 0.71)~\cite{Foo_2004_PhysRevLett}.
The temperature dependence of the HWHM for these two Na modes shows no anomalies upon warming up across $T_\text{c1}$~[Figs.~\ref{Fig5_fitting_parameters}(b) and (e)], but the HWHMs get significantly larger upon warming up to $T^*$. They cannot be described by the conventional anharmonic decay model (Appendix~\ref{Anharmonic_decay_model})~\cite{Klemens_PhysRev148,Cardona_PRB1984}. 
The large linewidth broadening for the two Na modes suggests that large phonon-phonon anharmonic interaction damps these two Na-phonon modes and makes them vanish near $T^*$~[Figs.~\ref{Fig5_fitting_parameters}(b) and (e)]. The disappearance of these two Na modes can also be seen from the temperature dependence of the integrated intensity, which tends to be zero close to $T^*$~[Figs.~\ref{Fig5_fitting_parameters}(c) and (f)].

\begin{figure}[!t] 
\begin{center}
\includegraphics[width=1\columnwidth]{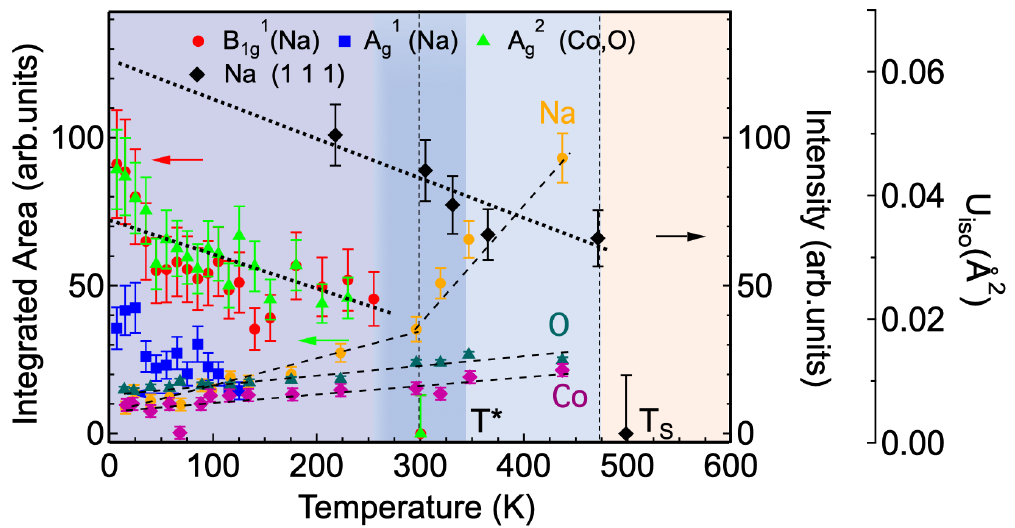}
\end{center}
\caption{\label{Fig6_xray}
The left axis shows the temperature dependence of the integrated intensity for the $B_{1g}^1$(Na) mode, $A^1_{g}$(Na) mode, and $A^2_{g}$(Co,O) mode of Na$_{0.5}$CoO$_2$. The \textcolor{black}{first} right axis shows the temperature dependence of 
 the (1~1~1) Bragg reflection in the orthorhombic $Pnmm$ setting adapted from Ref.~\cite{Argyriou_2007PhysRevB}. This reflection is a superlattice reflection with respect to the parent $P6_3/mmc$ crystal structure and arises due to Na ions zig-zag ordering. 
 \textcolor{black}{The data points labeled by ``Na'', ``O'' and ``Co'' are the $U_\text{iso}$ value (the second right axis) for Na, O and Co lattice adapted from Ref.~\cite{Argyriou_2007PhysRevB}, respectively.}
The thin dashed lines are guides for the eye, while the vertical dashed lines represent $T^*$ and $T_\text{S}$.
}   
\end{figure}

In Fig.~\ref{Fig6_xray}, we summarize the temperature dependence of the integrated intensity for the $B_{1g}^1$(Na), $A_g^1$(Na), and $A_g^2$(Co, O) modes, as well as the temperature dependence of the neutron powder diffraction (1~1~1) reflection in the orthorhombic $Pnmm$ setting adapted from Ref.~\cite{Argyriou_2007PhysRevB}. This (1~1~1) reflection is a superlattice reflection with respect to the parent $P6_3/mmc$ crystal structure \textcolor{black}{(absent in the parent $P6_3/mmc$ description)} that arises due to \textcolor{black}{long-range} Na ions zig-zag ordering. 
\textcolor{black}{
The temperature dependence of this Bragg peak's intensity tracks the structural zig-zag order.
}
The Na zig-zag order appears below $T_\text{S}\sim460$\,K, while the Na phonon mode emerges below a crossover temperature \textcolor{black}{$T^*\sim300\pm50$\,K} inferred from the $B_{1g}^1$(Na) mode.
For the $A_g^2$(Co,O) mode, while it is not a Na-dominant lattice vibration mode, it appears roughly at the same temperature $T^*$ as the $B_{1g}^1$(Na) mode, suggesting that the Na motion affects the nearby O vibration, which creates damping for the Co-O lattice vibration mode.
                   
Finally, we discuss the implications of the crossover temperature \textcolor{black}{$T^*\sim300\pm50$\,K} in Na$_{0.5}$CoO$_2$.  Above $T^*$, the Na-phonon mode shows large linewidth broadening, and becomes over-damped and invisible. Below $T^*$, the Na-phonon modes become underdamped, well-defined, and eventually resolution-limited.
Generally, a phonon mode appears abruptly after a first-order phase transition or gradually after a second-order phase transition for a solid sample. 
For Na$_{0.5}$CoO$_2$, the Na-phonon mode surprisingly appears at a much lower temperature, namely, 160\,K below the Na-zigzag structure phase transition $T_\text{S}$. This implies that the Na lattice becomes well-ordered, coherent, and \textcolor{black}{static} only below $T^*$. 
%\textcolor{red}{\sout{The crossover temperature $T^*$ implies a thermally-activated energy scale of about 25\,meV} for the Na ions to hop between neighboring zig-zag ordered Na potential wells.} 
Above $T^*$, the Na ions cannot be well bounded within a single Na potential well, and become dynamical and mobile, hopping along the zig-zag chain directions~\cite{Medarde_2013_PhysRevLett}.
The dynamical and mobile picture is consistent with the Rietveld analysis of the neutron powder diffraction data of Na$_{0.5}$CoO$_2$~\cite{Argyriou_2007PhysRevB}. The isotropic atomic displacement parameters \textcolor{black}{(mean-squared displacement entering the Debye–Waller factor)} U$_\text{iso}$ for the O and Co atoms are of the expected amplitude and show a linear behavior with temperature \textcolor{black}{(Fig.~\ref{Fig6_xray})}. In contrast, the U$_\text{iso}$ values for the Na atoms show a clear change in slope at $T^*$, and U$_\text{iso}$ becomes much larger above $T^*$\textcolor{black}{(Fig.~\ref{Fig6_xray})}. Such large values of U$_\text{iso}$ above $T^*$ are interpreted as dynamical motions of Na ions in between the CoO$_2$ layers~\cite{Argyriou_2007PhysRevB} (\textcolor{black}{Note that the U$_\text{iso}$ data is not a direct diffusion coefficient).}
\textcolor{black}{The mobile nature of the Na ions could create large damping for the collective motion of the Na lattice vibration modes; thus, they are not invisible above $T^*$.}

\textcolor{black}{
The timescale for the elastic neutron scattering process ranges from picoseconds to nanoseconds~\cite{karlsson_2015_proton}, while the timescale for Na hopping is microseconds at 300\,K~\cite{Mansson_2013_Physica_Scripta,Ohishi2023,tatara2025revisiting}. At lower temperatures, the Na hopping rate is much lower than at room temperature. Therefore, the Na hopping process is much slower than the elastic neutron scattering process. Elastic neutron scattering considers any motion occurring on a timescale significantly slower than the effective observation time of the instrument as ``static". This explains why elastic neutron scattering can detect the Na (1 1 1) superlattice Bragg peak in the presence of slow Na hopping~\cite{Argyriou_2007PhysRevB}.}

\textcolor{black}{
Raman scattering can indirectly detect slow Na dynamics/hopping by measuring how ion hopping affects the Na vibrational modes (phonons) of the crystal lattice. 
While Raman spectroscopy is technically an inelastic process that measures much higher energy scales (THz) than the hopping itself (MHz or lower), it ``sees" the Na dynamics/hopping through phonon line broadening and the loss of the integrated area. As Na ions begin to hop, the hopping increases the scattering of the Na phonons, creating a ``lifetime" effect. A shorter phonon lifetime results in a broadening of the Raman peaks, namely, the increase of the full width at half maximum. Meanwhile, it also leads to a weakening of the Na phonon modes. It is an open question how Na hopping affects the Raman phonon modes quantitatively using a microscopic Na-diffusion model in Na$_{0.5}$CoO$_2$.
Since Raman scattering is an indirect probe of slow Na dynamics/hopping, we cannot determine whether the crossover around 300\,K is a superionic transition in Na$_{0.5}$CoO$_2$~\cite{Weller_2009_PhysRevLett} based on our current Raman data.}

\textcolor{black}{
We note that the $A_g^1$(Na) mode at 157\,\cm-1 disappears at a temperature about 100\,K lower than the $B_{1g}^1$(Na) mode at 129\,\cm-1. 
On one hand, its intensity in the X’X’ scattering geometry is roughly half that of the 129\,\cm-1 mode~[Fig.~\ref{Fig4_T_dependence}(a)]. The weaker intensity makes it challenging to determine the exact disappearance temperature upon heating to 300\,K. Nonetheless, both Na-dominated modes exhibit similar weakening and broadening as the temperature rises to 300\,K, suggesting they may share the same coupling mechanism with the Na diffusion. On the other hand, the Na diffusion in Na$_{0.5}$CoO$_2$ is expected to be quasi-one-dimensional along the zig-zag chain direction, as revealed by neutron powder diffraction analysis of the orthorhombic phase of Na$_{0.7}$CoO$_2$~\cite{Medarde_2013_PhysRevLett}. 
Given that these two Na-dominated modes have parallel or perpendicular vibration directions relative to the Na zig-zag chain direction~[ $b$ axis shown in Fig.~\ref{Fig3_pattern}], they may couple to the Na diffusion with different strengths, resulting in different disappearance temperatures.
}

Our interpretation of a dynamical-to-\textcolor{black}{static} crossover involving mainly the motion of Na ions in Na$_{0.5}$CoO$_2$ is also consistent with a previous $^{23}$Na NMR study~\textcolor{black}{which has a timescale of microseconds to milliseconds}~\cite{Bobroff_2006_PhysRevLett}. At high temperature $T>210$\,K, a single and narrow quadrupolar peak is detected, while two Na crystallographic sites can only be resolved at lower temperatures. 
The single Na quadrupolar peak results from the occurrence of Na ionic diffusion at higher temperatures, which freezes at lower temperatures and gives rise to two Na crystallographic sites~\cite{Bobroff_2006_PhysRevLett}.

%Furthermore, the rapid rise of the spin-lattice relaxation rate $^{23}$Na $1/T_1T$ upon increasing temperature is also suggestive of increasing mobility and diffusion of Na ions above 210\,K, consistent with our results~\cite{Weller_2009_PhysRevLett,Lang_2008_PhysRevB.78.155116,Yokoi_2005_JPSJ}. 

%\sout{Compared with other Na contents in Na$_{x}$CoO$_2$ which is incommensurate with the Co-lattice, the sodium-ion self-diffusion coefficient for Na$_{0.5}$CoO$_2$ is an order of magnitude smaller based on the electrochemical measurements~\cite{Shu_2008_PhysRevB}. Although the mobility of the sodium ions increases upon warming up above $T^*$ for Na$_{0.5}$CoO$_2$, it is a relatively stable structure at room temperature and below $T^*$. }

%\textcolor{red}{\sout{Our results reconcile the NMR~\cite{Bobroff_2006_PhysRevLett} and electrochemical measurements~\cite{Shu_2008_PhysRevB} that the physics lies in the dynamical-to-localized crossover around $T^*\sim300$\,K in Na$_{0.5}$CoO$_2$.}}

The gradual freezing of the Na ions has an increasing impact from the Na-ion potential on the electronic properties of the CoO$_2$ layer, because Na ions become more and more confined to their mean crystallographic sites. This may explain why the in-plane magnetic susceptibility data $\chi(T)$ start to decrease below $T^*$~[Fig.~\ref{Fig1_structure}(d)].
%This deviation to a pure Pauli-constant behavior could be due to the strong correlations present in the 2D Co layer or any T depen- dence of the Fermi level density of states.
The well-defined and \textcolor{black}{static} Na-zigzag order below $T^*$ set the stage for the emergent charge order, the insulating state, and the antiferromagnetic orders in the CoO$_2$ layer of Na$_{0.5}$CoO$_2$~\cite{Roger_2007_nature,Zhang_2005_PhysRevB.71.153102,Zhou_2007_PhysRevLett.98.226402,Marianetti_2007_PhysRevLett,Choy_2007_PhysRevB.75.073103}.                                                                                                                                                            
                                                                                                                                                                                                                                                                                                                                                                            
\section{Conclusions}\label{Conclusions} 

In summary, we study the sodium-ion lattice dynamics in the Na zig-zag ordered cobaltate compound, Na$_{0.5}$CoO$_2$, using Raman spectroscopy and first-principles phonon calculations. 
Two previously unseen sodium phonon modes at 129 and 157\,\cm-1 are identified. Their mode frequencies are consistent with first-principles phonon calculations based on an orthorhombic unit cell. These modes emerge near room temperature (\textcolor{black}{$T^*\sim300\pm50$\,K}) with significant linewidth broadening, well below the established sodium zigzag ordering temperature ($T_\text{S}\sim460$\,K). These anomalies at $T^*$ are interpreted as a crossover from dynamical sodium motion to \textcolor{black}{static} states. We propose that this gradual freezing of sodium ions into a \textcolor{black}{static} zigzag configuration below $T^*$ provides the basis for the electronic and magnetic orders observed in the CoO$_2$ layer of Na$_{0.5}$CoO$_2$.

\begin{acknowledgments}
We acknowledge useful discussions with Professor Jianlin Luo.
This work was supported by the National Natural Science Foundation of China (Grants No.~12404548, No.~12574270, No.~12274186, and No.~12488201), 
the National Key Research and Development Program of China (Grant No. 2024YFA1408300, No.~2022YFA1403901).
The experimental work was supported by the Synergetic Extreme Condition User Facility (SECUF, \cite{SECUF}).                                                                                                                               
\end{acknowledgments}     
                           
%\cleardoublepage
             
\appendix

%\section{DFT band and phonon calculation} \label{band_phonon_dispersion}                                                                                                                             

%In this section, we discuss the DFT band and phonon calculation for Na$_{0.5}$CoO$_2$.

%The large discrepancies in the high-energy phonon modes between DFT and experiment may be related to the on-site Coulomb repulsion $U$ parameter used in the calculation. 
               
\section{Anharmonic phonon decay model}\label{Anharmonic_decay_model}    
In this section, we discuss the anharmonic phonon decay model.
We fit the temperature dependence of the phonon frequency and HWHM using the anharmonic phonon decay model~\cite{Klemens_PhysRev148,Cardona_PRB1984}: 
\begin{equation}
\label{eq_omega1}
\omega(T)=\omega_{0}-C_1\{1+ 2n(\Omega(T)/2) \},\\
\end{equation}
\begin{equation}
\label{eq_gamma1}
\Gamma(T)=\gamma_{0}+\gamma_1\{1+ 2n(\Omega(T)/2)\},
\end{equation}
%\begin{equation}
%\label{eq_omega2}
%\omega_2(T)=\omega_1(T)-C_2\left[1+ 3n(\Omega(T)/3)+3n^2(\Omega(T)/3)\right],
%\end{equation}
%\begin{equation}
%\label{eq_gamma2}
%\Gamma_2(T)=\Gamma_1(T)+\gamma_2\left[1+ 3n(\Omega(T)/3)+3n^2(\Omega(T)/3)\right],
%\end{equation}
where $\Omega(T)= \hbar \omega / k_BT$ and $n(x)=1/(e^x-1)$ is the Bose-Einstein distribution function. $\omega(T)$ and $\Gamma(T)$ involve mainly the three-phonon decay process, in which an optical phonon decays into two acoustic modes with equal energies and opposite momenta. 
%$\omega_2(T)$ and $\Gamma_2(T)$ involves additional four-phonon decay processes compared with $\omega_1(T)$ and $\Gamma_1(T)$. 

%\bibliography{biblio_long}

%merlin.mbs apsrev4-1.bst 2010-07-25 4.21a (PWD, AO, DPC) hacked
%Control: key (0)
%Control: author (0) dotless jnrlst
%Control: editor formatted (1) identically to author
%Control: production of article title (0) allowed
%Control: page (1) range
%Control: year (0) verbatim
%Control: production of eprint (0) enabled
%

\end{document}

\title{Supplemental Material for:\\ \textcolor{black}{Temperature-driven sodium ion {dynamical-to-static} crossover in the zig-zag ordered phase of Na$_{0.5}$CoO$_2$}}                           

\author{Shangfei~Wu$^\star$}
\email{wusf@baqis.ac.cn}
\affiliation{Beijing Academy of Quantum Information Sciences, Beijing 100193, China}

\author{Hengxin~Tan$^\star$}
\affiliation{Key Laboratory of Artificial Structures and Quantum Control (Ministry of Education), School of Physics and Astronomy, Shanghai Jiao Tong University, Shanghai 200240, China}

\author{Dong~Wu}
\affiliation{Beijing Academy of Quantum Information Sciences, Beijing 100193, China}

\author{Mingshu~Tan}
\affiliation{School of Physical Science and Technology, Lanzhou University, Lanzhou 730000, China}

\author{Xinyu~Zhou}
\affiliation{International Center for Quantum Materials, School of Physics, Peking University, Beijing 100871, China}
   
\author{Tianchen~Hu}
\affiliation{International Center for Quantum Materials, School of Physics, Peking University, Beijing 100871, China}
    
\author{Tao~Dong}
\affiliation{International Center for Quantum Materials, School of Physics, Peking University, Beijing 100871, China}

\author{Feng~Jin}
\affiliation{Beijing National Laboratory for Condensed Matter Physics, Institute of Physics, Chinese Academy of Sciences, Beijing 100190, China}

\author{Qingming~Zhang}
%\email{qmzhang@iphy.ac.cn}
\affiliation{Beijing National Laboratory for Condensed Matter Physics, Institute of Physics, Chinese Academy of Sciences, Beijing 100190, China}

\author{Nanlin~Wang}
%\email{nlwang@pku.edu.cn}
\affiliation{Beijing Academy of Quantum Information Sciences, Beijing 100193, China}
\affiliation{International Center for Quantum Materials, School of Physics, Peking University, Beijing 100871, China}
\affiliation{Tsung-Dao Lee Institute, Shanghai Jiao Tong University, Shanghai 200240, China}

\date{\today}               
                                                                                                                                                                                                              
%\begin{abstract}  
%     
%%\textcolor{black}{The sodium cobaltate Na$_x$CoO$_2$ has been extensively studied for its diverse functional properties along with its rich charge-spin-orbital ordered states. However, less is known about the sodium-ion lattice dynamics.}
%We employ polarization-resolved Raman spectroscopy combined with first-principles calculations to study the sodium-ion lattice dynamics in a sodium zig-zag ordered cobaltate compound Na$_{0.5}$CoO$_2$.
%We detect two sodium phonon modes for the first time, and their mode frequencies are consistent with first-principles phonon calculations based on an orthorhombic unit cell.
%We find that they appear below around \textcolor{black}{$T^*\sim300\pm50$\,K} with large linewidth broadening, much lower than the sodium zig-zag ordering temperature $T_\text{S}\sim460$\,K, and then narrow at lower temperatures. We interpret the sodium-phonon anomalies occurring at $T^*$ as a dynamical-to-\textcolor{black}{static} crossover involving mainly the motion of sodium ions. Our results suggest that the gradual freezing of the sodium ions and the well-defined \textcolor{black}{static} sodium-zigzag order below $T^*$ set the stage for the emergent electronic and magnetic orders in the CoO$_2$ layer of Na$_{0.5}$CoO$_2$.
                                                                       
%\end{abstract}                  
                                                                                                                                                                                      
                                                                                                                                                                                                                                                                                                                                                                                         
\maketitle

\begin{figure*}[!t] 
\begin{center}
\includegraphics[width=2\columnwidth]{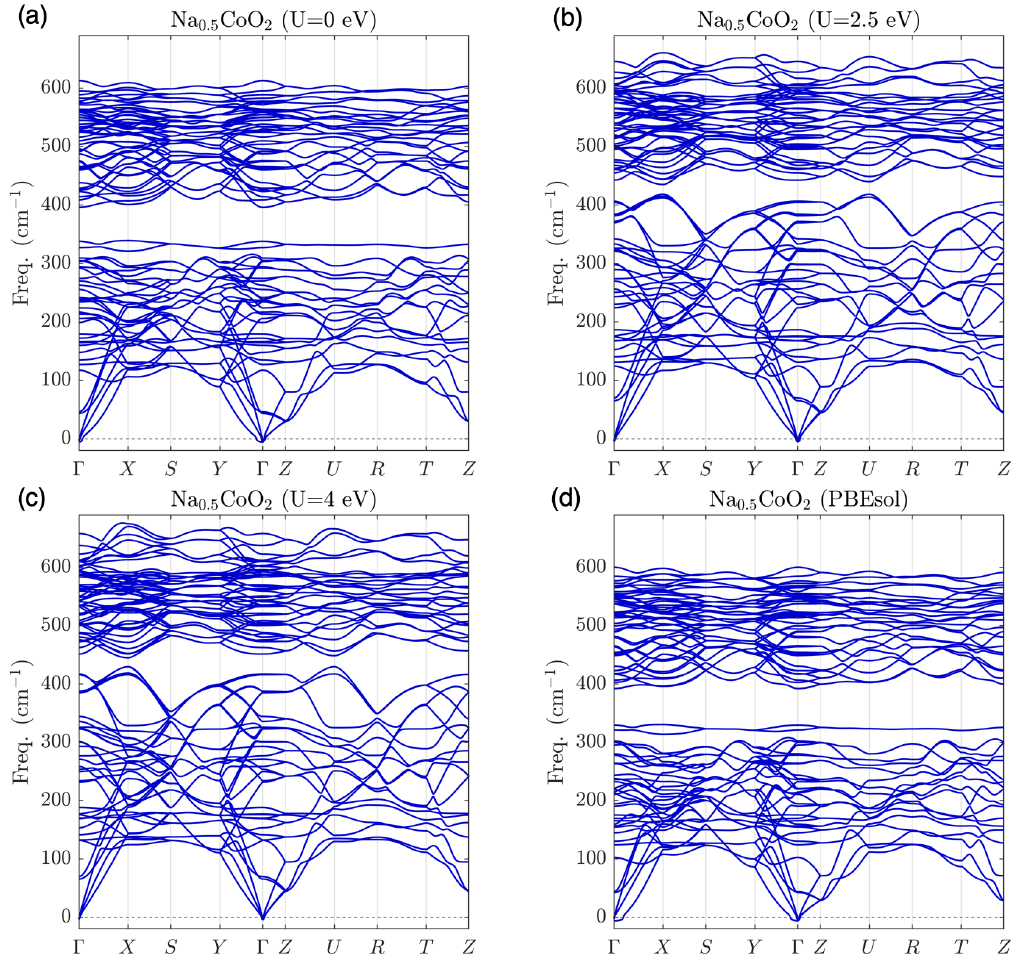}
\end{center}
\caption{\label{U_dependence} 
%\textcolor{cyan}{
Phonon spectrum under (a) Hubbard $U=0$\,eV, (b) $U=2.5$\,eV, (c) $U=4$\,eV, and (d) PBEsol without $U$ correction. The two lowest Raman-active phonon modes are: (a) 127 and 148\,\cm-1, (b) 131 and 153\,\cm-1, (c) 133 and 153\,\cm-1, (d) 124 and 145\,\cm-1. The experimental values are 127 and 157\,\cm-1.}
\end{figure*}                                                                                                          
                                        
\section{DFT band and phonon calculation} \label{band_phonon_dispersion}                                                                                                                             

%In this section, we discuss why we choose the Hubbard $U$ parameter $U=2.5$\,eV in the DFT calculation for Na$_{0.5}$CoO$_2$.

In this section, we discuss the DFT band and phonon calculation for Na$_{0.5}$CoO$_2$.

\textcolor{black}{
Previous studies have highlighted the strong sensitivity of the electronic structure of Na$_{0.5}$CoO$_2$ to the choice of the Hubbard $U$ parameter. For instance, Singh~\cite{Singh_2000PhysRevB} suggested an effective $U$ value of $5-8$\,eV for Na$_{0.5}$CoO$_2$. Li et al.~\cite{Li_2005_PhysRevB} reported a critical $U = 3.5$ eV, above which the ground state changes from ferromagnetic to antiferromagnetic. Lee and Pickett~\cite{Lee_2006_PhysRevLettDFT} proposed a much smaller critical value of $U \approx  0.5$\,eV, beyond which the largest local magnetic moments on Co become identical across different magnetic phases; Moreover, for $U \geq 2.5$\,eV, one Co sublattice becomes non-spin-polarized while the other exhibits strong spin polarization. In addition, Lee et al.~\cite{Lee_2004_PhysRevB.70.045104} identified a critical $U \approx 3$\,eV above which charge disproportionation occurs in related Na$_x$CoO$_2$ compounds. These results collectively indicate that no single Hubbard $U$ value can reliably capture all physical properties of Na$_{0.5}$CoO$_2$, reflecting its potential strong correlation effects.
}

\begin{figure*}[!t] 
\begin{center}
\includegraphics[width=2\columnwidth]{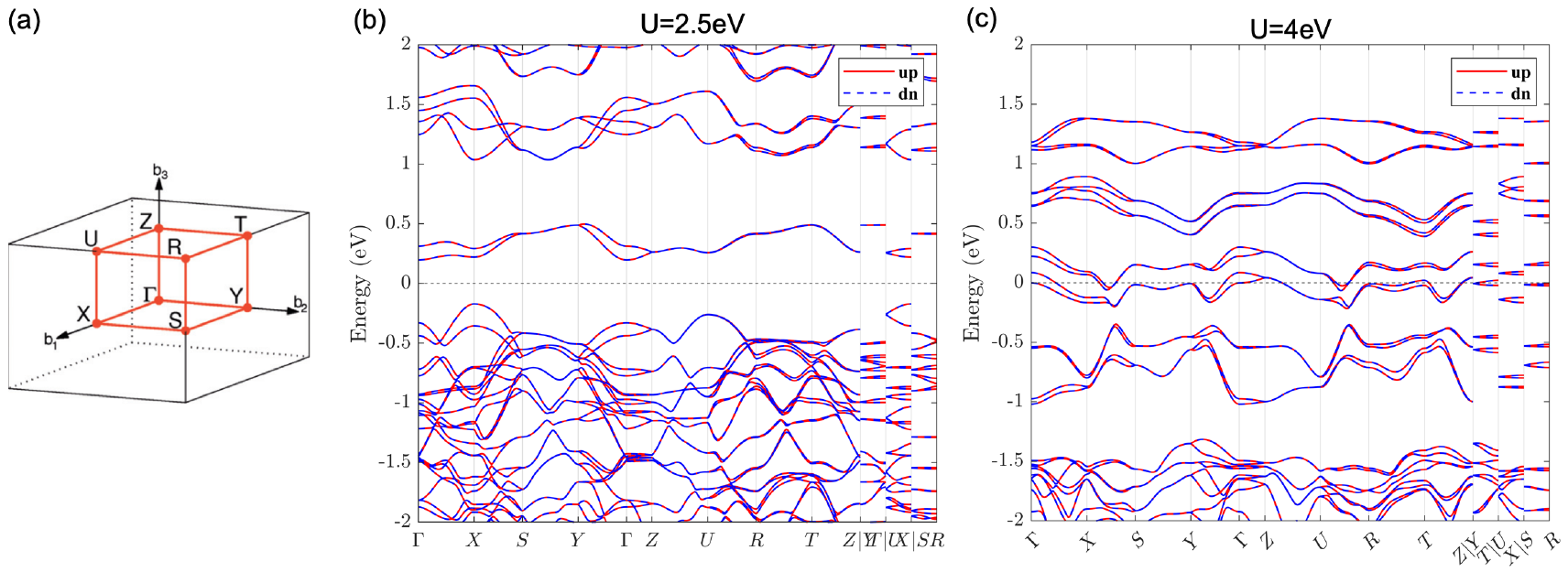}
\end{center}
\caption{\label{U=4eV} 
%\textcolor{cyan}{
(a) Orthorhombic Brillouin zone notations for the special $k$-points.
(b)Non-relativistic band structure of Na$_{0.5}$CoO$_2$ with Hubbard $U = 2.5$\,eV for Co 3$d$ electrons. The two spin channels are represented by ‘up’ and ‘dn’, respectively.
(c) Same as (b) but for  $U = 4$\,eV.
 }
\end{figure*}

\textcolor{black}{
We performed calculations with Hubbard $U = 0,~2.5, \text{and}~4$\,eV, as well as using the PBEsol exchange-correlation functional. The resulting phonon dispersions are shown in Fig.~\ref{U_dependence}. The two lowest Raman-active phonon modes are 127 and 148\,\cm-1 ($U=0$\,eV), 131 and 153\,\cm-1 ($U=2.5$\,eV), 133 and 153\,\cm-1 ($U=4$\, eV), and 124 and 145\,\cm-1 (PBEsol), compared to the experimental values of 127 and 157\,\cm-1. These results indicate that low-frequency Na modes with $U$ correction agree better with experiment, while the high-frequency oxygen-dominated modes show clear deviations across all calculations. This confirms that no single $U$ value can fully describe all phonon properties, consistent with previous studies~\cite{Singh_2000PhysRevB,Li_2005_PhysRevB,Lee_2006_PhysRevLettDFT,Lee_2004_PhysRevB.70.045104}. 
Thus, we chose $U = 2.5$\,eV to match our low-frequency Raman experiments. Moreover, this $U$ value reproduces an insulating electronic structure for Na$_{0.5}$CoO$_2$ [see Fig.~\ref{U=4eV}(b)], whereas the $U = 4$\,eV calculation yields a metallic one [see Fig.~\ref{U=4eV}(c)]. This choice of $U$ value also reproduces that the stripe antiferromagnetic configuration is energetically favored for Na$_{0.5}$CoO$_2$.}
   
\textcolor{black}{
Regarding the numerical uncertainty, we performed additional phonon calculations with (i) a varied cutoff energy (520\,eV), (ii) a varied displacement amplitude (0.005\,\AA), and (iii) a varied supercell ($3\times3\times2$). Results are shown in Fig.~\ref{numerical_uncertainty} below. The lowest Raman active phonon modes are 134 and 153\,\cm-1 (i), 134 and 154 \,\cm-1 (ii), 130 and 150\,\cm-1 (iii). These results fall within a reasonable margin of error.} 
\textcolor{black}{For the high-energy modes, the calculated phonon frequencies with cutoff energies of 400\,eV and 520\,eV show little differences (Table~\ref{phonon_modes_SI1}). We also tried the vdW correction (DFT-D3 method of Grimme with zero-damping function) to the calculations. Indeed, the out-of-plane lattice constant is slightly decreased from 10.61 to 10.43~\AA, while the in-plane lattice constants remain largely unchanged. However, this revision does not affect the phonon properties much. For example, the vdW correction leads to negligible changes to the dispersion as shown in Fig.~\ref{vdW_correction}, especially at the low frequency region. Thus, it is reasonable to ignore the vdW correction in our calculations.}

\begin{figure*}[!t] 
\begin{center}
\includegraphics[width=2\columnwidth]{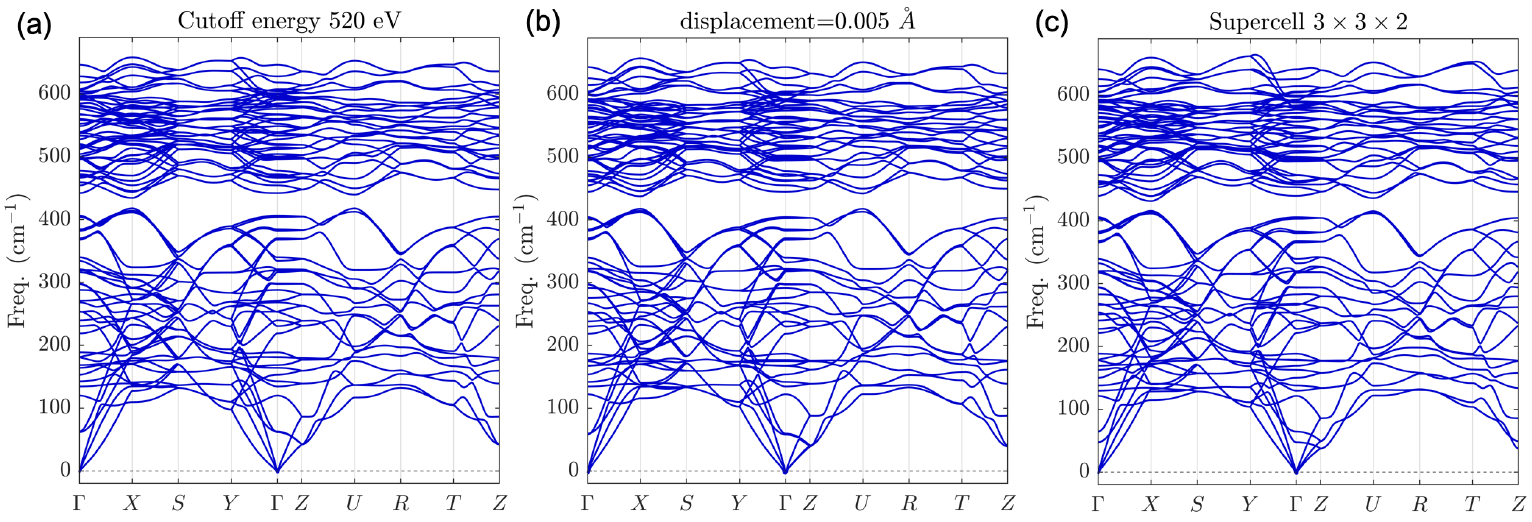}
\end{center}
\caption{\label{numerical_uncertainty} 
%\textcolor{cyan}{
Phonon dispersions of Na$_{0.5}$CoO$_2$ at $U=2.5$\,eV. Left panel: the cutoff energy for planewave basis set is 520\,eV. Middle panel: distortion displacement is 0.005\,\AA. Right panel: supercell size $3\times3\times2$. The lowest Raman active phonon modes are 134.3 and 153.0\,\cm-1 (left), 134.5 and 153.7\,\cm-1 (middle), 129.7 and 150.1\,\cm-1 (right). }
\end{figure*}

 \begin{figure}[t] 
\includegraphics[width=\columnwidth]{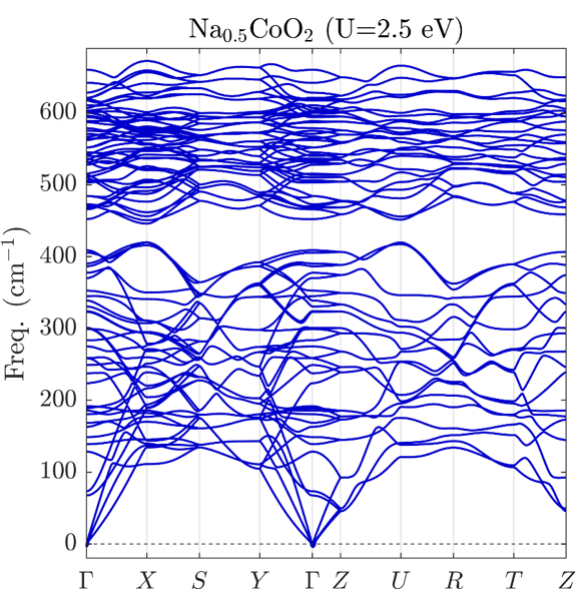}
\caption{\label{vdW_correction} 
Phonon dispersion of Na$_{0.5}$CoO$_2$ under $U=2.5$\,eV. Notice that the vdW correction is included in the structure relaxation.
 }
\end{figure}

In Fig.~\ref{U=4eV}(b), we show the calculated band structure for Na$_{0.5}$CoO$_2$ without spin-orbit coupling. The on-site Coulomb interaction $U=2.5$\,eV is used in the calculation.
The number of valence bands in Fig.~\ref{U=4eV}(b) is four times the bands shown in Ref.~\cite{Singh_2000PhysRevB}, because our calculation is based on the orthorhombic supercell with space group $Pnmm$ (No.~59) (point group:~$D_{2h}$), while a hexagonal structure with space group $P6_322$ was used in Ref.~\cite{Singh_2000PhysRevB}. 
The unit cell in the orthorhombic phase is four times the hexagonal unit cell [Fig.~1(c) of the main text].
%The same is true for the phonon dispersions [Fig.~\ref{band_phonon}(b)] compared with Ref.~\cite{Li_2004_PhysRevB}.

%\textcolor{black}{There have not to date been detailed photoemission studies of Na$_{0.5}$CoO$_2$ but trends for other octahedrally coordinated 3d transition metal compounds (i.e., U estimated from photoemission satellite positions as the atomic number and ionicity are varied) would suggest an effective on-site Hubbard U of 5-8 eV.}  
         
In Fig.~\ref{U_dependence}(b), we show the calculated DFT phonon dispersion with $U=2.5$\,eV. The frequencies of the Raman-active modes in the Brillouin-zone center are shown in Table~II of the main text.
Overall, the mode frequencies of the experimentally observed phonons are consistent with the DFT phonon calculations. Especially, the low-energy modes below 350\,\cm-1 are in good accordance with DFT phonon calculations. For the high-energy modes above 350\,\cm-1, which are mainly O-related vibration modes, \textcolor{black}{the calculated phonon frequencies are, on average, about 30\,\cm-1 larger than the experimental ones~[Table~II of the main text]}. 

%\begin{figure*}[!t] 
%\begin{center}
%\includegraphics[width=2\columnwidth]{band_phonon}
%\end{center}
%\caption{\label{band_phonon} 
%%\textcolor{cyan}{
%(a) Orthorhombic Brillouin zone notations for the special $k$-points.
%(b) Band structure for the orthorhombic Na$_{0.5}$CoO$_2$ without spin-orbit coupling.
%(c) Phonon dispersion calculated with $U=2.5$\,eV of the orthorhombic structure.}
%%}
%\end{figure*}  

\textcolor{black}{
Finally, we discuss the possible sources of the large mismatch between the calculated and experimental phonon frequencies for the high energy modes above 350\,\cm-1. We consider the discrepancy to arise from a combination of the following factors:
First, the properties of Na$_{0.5}$CoO$_2$ are known to be sensitive to the choice of $U$ values~\cite{Singh_2000PhysRevB,Li_2005_PhysRevB,Lee_2006_PhysRevLettDFT,Lee_2004_PhysRevB.70.045104}, and it is generally difficult to identify a single $U$ value that simultaneously reproduces all experimental observables. 
We have adopted $U = 2.5$\,eV to provide a reasonable description of the low-frequency phonon modes that are central to our analysis. 
This choice, however, may contribute to a less satisfactory agreement in the high-frequency region.
Second, it is well established that standard exchange–correlation functionals can lead to small but non-negligible deviations in lattice dynamics (for instance, LDA typically underestimates while GGA overestimates the lattice constant), even for weakly correlated materials. 
Third, our calculations are performed within the harmonic approximation. Although anharmonic effects are expected to be limited at low temperatures, they may still influence the high-frequency modes to some extent.
Overall, while the agreement between DFT and experiment is not uniform across all modes, the level of discrepancy remains within a few meV (30\,\cm-1 $\sim$ 3.7 meV), which is in the energy accuracy limit of common DFT calculations. The main physical conclusions in the manuscript are not affected. %Nevertheless, we revise the manuscript to include a more explicit discussion of the possible sources of discrepancy, as suggested by the referee.
}

\section{Lattice vibration patterns for $A_g$ and $B_{1g}$ modes in Na$_{0.5}$CoO$_2$}\label{Lattice_vibration_patterns}    
      
\begin{figure*}[!t] 
\begin{center}
\includegraphics[width=2\columnwidth]{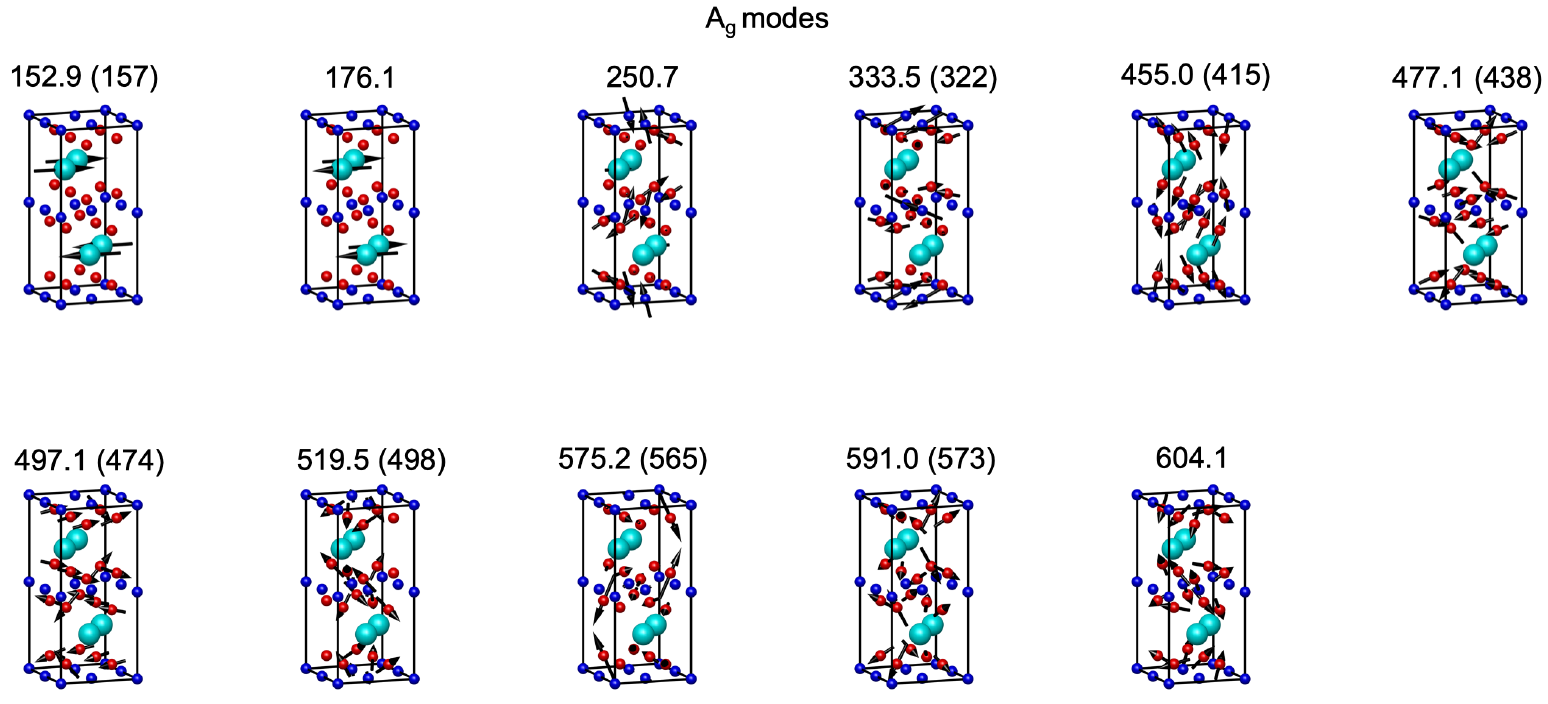}
\end{center}
\caption{\label{Ag_modes} 
%\textcolor{cyan}{
Lattice vibration patterns for the $A_g$ Raman-active modes in Na$_{0.5}$CoO$_2$. The numeric values on top of each figure are the calculated phonon mode frequencies. The corresponding experimental phonon frequencies are in parentheses.
The units are in \cm-1. 
}
%}
\end{figure*}

\begin{figure*}[!t] 
\begin{center}
\includegraphics[width=1.6\columnwidth]{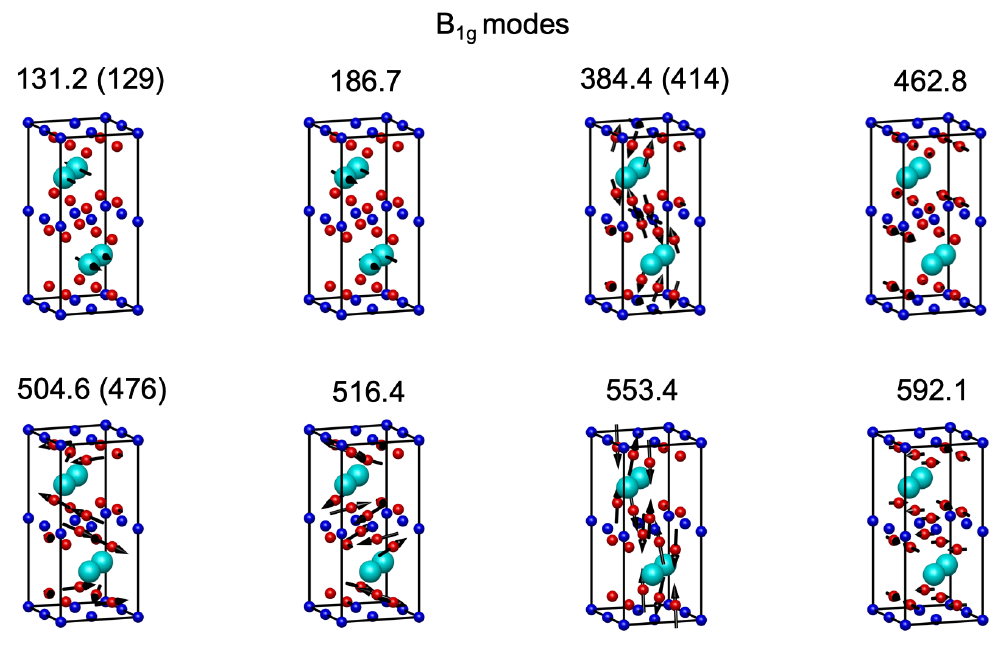}
\end{center}
\caption{\label{B1g_modes} 
%\textcolor{cyan}{
Lattice vibration patterns for the $B_{1g}$ Raman-active modes in Na$_{0.5}$CoO$_2$. The numeric values on top of each figure are the calculated phonon mode frequencies. The corresponding experimental phonon frequencies are in parentheses.
The units are in \cm-1. 
}
%}
\end{figure*}

In this section, we show the lattice vibration patterns for the Raman-active modes in Na$_{0.5}$CoO$_2$. We focus on the $A_g$ and $B_{1g}$ modes that are accessible in the $ab$-plane measurement.  Fig.~\ref{Ag_modes} shows the lattice vibration patterns for the $A_g$ modes, while Fig.~\ref{B1g_modes} shows the lattice vibration patterns for the $B_{1g}$ modes. 

Based on Fig.~\ref{Ag_modes} and Fig.~\ref{B1g_modes}, we find that the first four low-energy modes ($B_{1g}$ mode at 131.2\,\cm-1, $A_g$ modes at 152.9\,\cm-1 and 176.1\,\cm-1, $B_{1g}$ mode at 186.7\,\cm-1) have dominantly Na-related lattice vibration patterns. The high-energy modes above 350\,\cm-1 have dominantly O-related lattice vibration patterns. At intermediate energies, the lattice vibration patterns are mainly Co and O related vibrations.

\newpage

\begin{table*}
\caption{
\label{phonon_modes_SI1} 
Calculated phonon frequencies at the Brillouin zone center for Na$_{0.5}$CoO$_2$ by DFT with cuttoff energy 400 and 520\,eV. All the units are in \cm-1.}
\begin{ruledtabular}
\begin{tabular}{cc|cc}
Symmetry& Cutoff Energy 400\,eV &Symmetry &Cutoff Energy 520\,eV \\
\hline 
 $A_u$ & 65.3 &$A_u$  &61.7 \\
 $B_{1u}$ & 70.0 & $B_{1u}$ &62.8 \\
 $B_{2u}$ & 124.5 & $B_{2u}$  &119.9 \\
 $B_{1g}$ & 131.2 &  $B_{1g}$ &134.3 \\
 $B_{3u}$ & 140.2 & $B_{3u}$ &142.5  \\
 $A_g$ & 152.9 & $A_g$ & 153.0 \\
 $B_{2u}$ & 167.7 & $B_{2u}$ & 164.2\\
 $B_{3u}$ & 168.3 &$B_{3u}$  & 168.5 \\
 $B_{2u}$ & 174.8 & $A_g$ &180.8  \\
 $A_g$ & 176.1 & $B_{2u}$ & 181.2 \\
 $B_{1g}$ & 186.7 & $B_{1g}$& 188.9\\
 $A_u$ & 224.2 & $A_u$ &218.8 \\
 $B_{3g}$ & 230.2 & $B_{3g}$  &230.6  \\
 $B_{3u}$ & 243.1 &$B_{3u}$ &239.5  \\
 $A_g$ & 250.7 & $A_g$ & 253.3 \\
 $B_{1u}$ & 256.6 & $B_{1u}$ &253.7  \\
 $B_{2u}$ & 270.0 & $B_{2u}$ &268.3 \\
 $B_{1u}$ & 273.9 &$B_{1u}$   &271.0 \\
 $B_{3g}$ & 289.4 & $B_{3g}$  &288.1 \\
 $A_u$ & 298.3 & $A_u$ &297.3 \\
 $B_{3u}$ & 298.8 &$B_{3u}$  &297.8 \\
 $B_{3g}$ & 321.1 & $B_{3g}$ &316.4 \\
 $A_{u}$ & 323.5 & $B_{1u}$ & 319.3 \\
 $B_{3u}$ & 324.1 & $A_u$  &320.1  \\
 $B_{1u}$ & 325.8 &$B_{3u} $   & 320.9\\
 $A_{g}$ & 333.5 & $A_g$  &334.4  \\
 $B_{3g}$ & 342.3 & $B_{3g}$  & 339.9 \\
 $B_{2u}$ & 369.5 &$B_{2u}$   &368.1  \\
 $B_{1u}$ & 372.0 & $B_{1u}$ &370.3 \\
 $B_{2g}$ & 381.6 & $B_{2g}$   & 382.1\\
 $B_{1g}$ & 384.4 & $B_{1g}$   &384.6 \\
 $B_{1u}$ & 404.7 & $B_{1u}$   & 404.0\\
 $B_{2u}$ & 406.8 & $B_{2u}$   & 406.1\\
 $B_{3g}$ & 442.6 &  $B_{3g}$& 442.1  \\
 $A_{g}$ & 455.0 & $A_g$   &454.6 \\
 $A_{u}$ & 460.8 &  $A_u$  &459.6 \\
 $B_{2g}$ & 462.0 & $B_{2g}$   &466.6 \\
 $B_{1g}$ & 462.8 & $B_{1g}$  &467.5 \\
 $B_{3g}$ & 467.9 & $B_{3g}$   &469.4 \\
 $B_{3u}$ & 473.5 & $B_{3u}$    &472.2 \\
 $A_{g}$ & 477.1 & $A_g$   &478.1 \\
 $B_{3g}$ & 495.3 & $B_{3g}$   &496.0 \\
 $A_{g}$ & 497.1 &  $A_g$  &497.7 \\
 $A_{u}$ & 499.1 & $A_u$    &497.9 \\
 $B_{2g}$ & 500.3 &  $B_{2g}$  &500.2 \\
 $B_{3u}$ & 501.8 &  $B_{3u}$  &500.4 \\
 $B_{1g}$ & 504.6 & $B_{1g}$  &504.2 \\
 $B_{2g}$ & 513.9 & $B_{2g}$   &512.6 \\
 $B_{1g}$  & 516.4 &  $B_{1g}$ & 515.2\\
 $A_{g}$ & 519.9 &  $B_{3g}$  & 518.6\\
 $B_{3g}$ & 519.7 &  $A_g$  &520.3 \\
 $B_{1u}$ & 526.6 & $B_{1u}$  & 527.8 \\
 $A_{u}$ & 532.0 & $A_u$   &533.7 \\
 $B_{2u}$ & 536.3 & $B_{2u}$  & 537.7 \\
 $B_{3u}$ & 536.4 &$B_{3u}$   & 538.1 \\
 $B_{1u}$ & 539.8 & $B_{1u}$   & 539.6\\
 $B_{2g}$ & 539.9 & $B_{2g}$   &539.6 \\
 $B_{2u}$ & 549.1 & $B_{2u}$   & 548.8\\
 $B_{1g}$ & 553.4 & $B_{1g}$   &552.6 \\
 $A_{u}$ & 557.9 &  $B_{1u}$  & 558.3\\
 $B_{1u}$ & 557.6 & $A_u$   & 560.3 \\
\end{tabular}
\end{ruledtabular}
\end{table*}

\begin{table*}
\ContinuedFloat
\caption{
\label{phonon_modes_SI2} 
Calculated phonon frequencies at the Brillouin zone center for Na$_{0.5}$CoO$_2$ by DFT with cuttoff energy 400 and 520\,eV. All the units are in \cm-1. (Continued)}
\begin{ruledtabular}
\begin{tabular}{cc|cc}
Symmetry& Cutoff Energy 400\,eV & Symmetry &Cutoff Energy 520\,eV \\
\hline 
 $B_{3u}$ & 564.9 & $B_{2u}$  &566.6  \\
 $B_{2u}$ & 565.8 &  $B_{3u}$&567.7   \\
 $A_{g}$ & 575.2 &  $B_{3g}$  &576.4 \\
 $B_{3g}$ & 575.70 & $A_u$  &579.9  \\
 $A_{u}$ & 578.0 & $B_{3u}$   & 581.3\\
 $B_{3u}$ & 581.3 &  $A_g$ &581.9  \\
 $B_{1u}$ & 584.9 &  $B_{1u}$ & 584.9 \\
 $B_{2g}$ & 589.7 & $B_{2u}$  &591.0 \\
 $B_{2u} $& 590.6 & $B_{2g}$  &593.4  \\
 $A_{g}$ & 591.0 & $A_g$  & 594.1 \\
 $B_{1g}$ & 592.1 & $B_{1u}$  & 594.3 \\
 $B_{1u}$ & 594.5 &  $B_{1g}$  &595.9 \\
 $B_{3g}$ & 596.6& $B_{3g}$ & 599.0\\
 $B_{3g}$ & 599.9 & $B_{3g}$  &601.1 \\
 $A_{g} $& 604.1 &  $A_g$ &  604.6\\
 $B_{1u}$ & 605.3 & $B_{1u}$ & 606.0 \\
 $B_{2u}$ & 607.1 & $B_{2u}$ & 607.4 \\
 $B_{2u}$ & 617.1 & $B_{2u}$&  618.4\\
 $B_{1u}$  & 627.1 & $B_{1u}$ & 628.0 \\
 $B_{2u}$  & 645.6 &$B_{2u}$  & 646.7\\
\end{tabular}
\end{ruledtabular}
\end{table*}

%\bibliography{biblio_long}
%merlin.mbs apsrev4-1.bst 2010-07-25 4.21a (PWD, AO, DPC) hacked
%Control: key (0)
%Control: author (0) dotless jnrlst
%Control: editor formatted (1) identically to author
%Control: production of article title (0) allowed
%Control: page (1) range
%Control: year (0) verbatim
%Control: production of eprint (0) enabled
%